\documentclass[11pt]{article}
\usepackage{graphicx,setspace,verbatim} % Add graphics capabilities, spacing, verbatim
\usepackage[colorlinks=true,linkcolor=blue]{hyperref} % Hyperlink capabilities
\usepackage[margin = 0.75in]{geometry}
\usepackage{amsmath}
\usepackage{apalike}
\long\def\comment#1{}

\newcommand{\marg}{\hat{\theta}_{\mathrm{MPELE}}}
\newcommand{\margNR}{\hat{\theta}_{\mathrm{MELE}}} % no regulation
\newcommand{\mle}{\hat{\theta}_{MLE}}
\newcommand{\map}{\hat{\theta}_{MAP}}
 
\newcommand{\epec}[1]{\mathbf{E}\Big[ #1 \Big]}

\title{\begin{singlespace} Fast inference in generalized linear models
    via expected log-likelihoods \end{singlespace}}
\date{}

\author{{Alexandro D. Ramirez}\/\raisebox{4pt}{\scriptsize$\;$1,*}, 
{Liam Paninski}\/\raisebox{4pt}{\scriptsize$\;$2}\\\\
\raisebox{4pt}{\scriptsize 1$\;$}Weill Cornell Medical College, NY. NY, U.S.A\\
\raisebox{4pt}{\scriptsize *$\;$}alr2038@med.cornell.edu\\
\raisebox{4pt}{\scriptsize 2$\;$}Columbia University Department of Statistics, \\
Center for Theoretical Neuroscience, Grossman Center for the Statistics of Mind, \\
 Kavli Institute for Brain Science, NY. NY, U.S.A}

\begin{document}
\maketitle
\begin{abstract}
Generalized linear models play an essential role in a wide variety of
statistical applications.  This paper discusses an approximation of
the likelihood in these models that can greatly facilitate
computation.  The basic idea is to replace a sum that appears in the
exact log-likelihood by an expectation over the model covariates; the
resulting ``expected log-likelihood'' can in many cases be computed
significantly faster than the exact log-likelihood.  In many
neuroscience experiments the distribution over model covariates is
controlled by the experimenter and the expected log-likelihood
approximation becomes particularly useful; for example, estimators
based on maximizing this expected log-likelihood (or a penalized
version thereof) can often be obtained with orders of magnitude
computational savings compared to the exact maximum likelihood
estimators.  A risk analysis establishes that these maximum EL
estimators often come with little cost in accuracy (and in some cases
even improved accuracy) compared to standard maximum likelihood
estimates.  Finally, we find that these methods can significantly
decrease the computation time of marginal likelihood calculations for
model selection and of Markov chain Monte Carlo methods for sampling
from the posterior parameter distribution.  We illustrate our results
by applying these methods to a computationally-challenging dataset of
neural spike trains obtained via large-scale multi-electrode
recordings in the primate retina.
\end{abstract}

\section{Introduction}
Systems neuroscience has experienced impressive technological
development over the last decade.  For example, ongoing improvements
in multi-electrode recording \cite{brown04a,field10a,stevenson11a} and
imaging techniques \cite{cossart03a,ohki05a,lutcke10a} have made it
possible to observe the activity of hundreds or even thousands of
neurons simultaneously.  To fully realize the potential of these new
high-throughput recording techniques, it will be necessary to develop
analytical methods that scale well to large neural population sizes.
The need for efficient computational methods is especially pressing in
the context of on-line, closed-loop experiments
\cite{Donoghue02,Santhanam06,Lewi08}.

Our goal in this paper is to develop scalable methods for neural spike
train analysis based on a generalized linear model (GLM) framework
\cite{mccullagh89a}.  This model class has proven useful for
quantifying the relationship between neural responses and external
stimuli or behavior
\cite{brillinger88a,paninski04b,truccolo05a,pillow08a,truccolo10a},
and has been applied successfully in a wide variety of brain areas;
see, e.g., \cite{vidne11a} for a recent review (Of course, GLMs are
well-established as a fundamental tool in applied statistics more
generally.)  GLMs offer a convenient likelihood-based approach for
predicting responses to novel stimuli and for neuronal decoding
\cite{PAN06c}, but computations involving the likelihood can become
challenging if the stimulus is very high-dimensional or if many
neurons are observed.

The key idea presented here is that the GLM log-likelihood can be
approximated cheaply in many cases by exploiting the law of large
numbers: we replace an expensive sum that appears in the exact
log-likelihood (involving functions of the parameters and observed
covariates) by its expectation to obtain an approximate ``expected
log-likelihood'' (EL) (a phrase coined by Park and Pillow in
\cite{park11}).  Computing this expectation requires knowledge, or at
least an approximation, of the covariate distribution.  In many
neuroscience experiments the covariates correspond to stimuli which
are under the control of the experimenter, and therefore the stimulus
distribution (or least some moments of this distribution) may be
considered known a priori, making the required expectations
analytically or numerically tractable.  The resulting EL approximation
can often be computed significantly more quickly than the exact
log-likelihood.  This approximation has been exploited previously in
some special cases (e.g., Gaussian process regression
\cite{sollich05a,rasmussen05a} and maximum likelihood estimation of a
Poisson regression model
\cite{paninski04b,field10a,park11,sadeghi12a}).  We generalize the
basic idea behind the EL from the specific models where it has been
applied previously to all GLMs in canonical form and discuss the
associated computational savings.  We then examine a number of novel
applications of the EL towards parameter estimation, marginal
likelihood calculations, and Monte Carlo sampling from the posterior
parameter distribution.

\section{Results}

\subsection{Generalized linear models}

Consider a vector of observed responses, $r = (r_1, ..., r_N)$,
resulting from $N$ presentations of a $p$ dimensional stimulus vector,
$x_{i}$ (for $i=1, ...,N$).  Under a GLM, with model parameters
$\theta$, the likelihood for $r$ is chosen from an exponential family
of distributions \cite{lehmann98a}.  If we model the observations as
conditionally independent given $x$ (an assumption we will later
relax), we can write the log-likelihood for $r$ as
\begin{eqnarray}
\label{logl1}
	 L(\theta) \equiv \log p(r|\theta, \{ x_i\}) = \sum_{n=1}^N
         \frac{1}{c(\phi)} \Big( a(x_n^T \theta) r_n - G(x_n^T\theta)
         \Big) + \mathrm{const(\theta)},
\end{eqnarray}
for some functions $a( ), G( )$, and $c( )$, with $\phi$ an auxiliary
parameter \cite{mccullagh89a}, and where we have written terms that
are constant with respect to $\theta$ as $\mathrm{const(\theta)}$.
For the rest of the paper we will consider the scale factor $c(\phi)$
to be known and for convenience we will set it to one.  In addition,
we will specialize to the ``canonical'' case that $a(x_n^T \theta) =
x_n^T \theta$, i.e., $a(.)$ is the identity function.  With these
choices, we see that the GLM log-likelihood is the sum of a linear and
non-linear function of $\theta$,
\begin{eqnarray}
\label{logl}
	 L(\theta) &=&  \sum_{n=1}^N (x_n^T \theta) r_n -   G(x_n^T\theta) + \mathrm{const(\theta)}.
\end{eqnarray}
This expression will be the jumping-off point for the EL
approximation.  However, first it is useful to review a few familiar
examples of this GLM form.

First consider the standard linear regression model, in which the
observations $r$ are normally distributed with mean given by the inner
product of the parameter vector $\theta$ and stimulus vector $x$.  The
log-likelihood for $r$ is then
\begin{eqnarray}
       L(\theta) &=& \sum_{n=1}^N -\frac{ (r_n - x_n^T\theta)^2}{2
         \sigma^2}+ \mathrm{const(\theta)}\\ &\propto& \sum_{n=1}^N
       (x_n^T\theta) r_n - \frac{1}{2}(x_n^T\theta)^2+
       \mathrm{const(\theta)},
\end{eqnarray}
where for clarity in the second line we have suppressed the scale
factor set by the noise variance $\sigma^2$.  The non-linear function
$G(.)$ in this case is seen to be proportional to
$\frac{1}{2}(x_n^T\theta)^2$.

As another example, in the standard Poisson regression model,
responses are distributed by an inhomogeneous Poisson process whose
rate is given by the exponential of the inner product of $\theta$ and
$x$.  (In the neuroscience literature this model is often referred to
as a linear-nonlinear-Poisson (LNP) model \cite{Simoncelli04}.)  If we
discretize time so that $r_n$ denotes the number of events (e.g., in
the neuroscience setting the number of emitted spikes) in time bin
$n$, the log-likelihood is
\begin{eqnarray}
       L(\theta) &=& \sum_{n=1}^N \log \left( \exp \Big(-\exp(x_n^T
           \theta) \Big) \frac{\Big(\exp(x_n^T \theta) \Big)^{r_n}} { r_n!} \right)\\
       		    &=&  \sum_{n=1}^N (x_n^T \theta) r_n  - \exp(x_n^T
       \theta)+ \mathrm{const(\theta)}.
\end{eqnarray}
In this case we see that $G(.) = \exp(.)$.

As a final example, consider the case where responses are distributed
according to a binary logistic regression model, so that $r_n$ only
takes two values, say $0$ or $1$, with $p_n \equiv p(r_n =1 |x_n^T,
\theta)$ defined according to the canonical ``logit'' link function 
\begin{eqnarray}
  \log \left( \frac {p_n} {1-p_n} \right) = x_n^T\theta.
\end{eqnarray}
Here the log-likelihood is
\begin{eqnarray}
      L(\theta) &=& \sum_{n=1}^N \log{\Big( p_n^{r_n}(1-p_n)^{1-r_n}
        \Big)}\\ &=& \sum_{n=1}^N (x_n^T\theta) r_n +
      \log{(1-p_n)}\\ &=& \sum_{n=1}^N (x_n^T\theta) r_n -
      \log{\Big(1+\exp\Big(x_n^T \theta\Big)\Big)},
\end{eqnarray}
so $G(.) = \log{\Big(1+\exp(.)\Big)}$.

\subsection{The computational advantage of using expected log-likelihoods over log-likelihoods in a GLM}
Now let's examine eq.~(\ref{logl}) more closely.  For large values of
$N$ and $p$ there is a significant difference in the computational
cost between the linear and non-linear terms in this expression.
Because we can trivially rearrange the linear term as $\sum_{n=1}^N
(x_n^T \theta) r_n = (\sum_{n=1}^N x_n^T r_n)\theta$, its computation
only requires a single evaluation of the weighted sum over vectors $x$  $\sum_{n=1}^N (x_n^Tr_n)$, no matter how many times the log-likelihood is evaluated.
(Remember that the simple linear structure of the first term is a
special feature of the canonical link function; our results below
depend on this canonical assumption.)  More precisely, if we evaluate
the log-likelihood $K$ times, the number of operations to compute the
linear term is $O(Np+Kp)$; computing the non-linear sum, in contrast,
requires $O(KNp)$ operations in general.  Therefore, the main burden
in evaluating the log-likelihood is in the computation of the
non-linear term.  The EL, denoted by $\tilde{L}(\theta)$, is an
approximation to the log-likelihood that can alleviate the
computational cost of the non-linear term.  We invoke the law of large
numbers to approximate the sum over the non-linearity in equation
\ref{logl} by its expectation
\cite{paninski04b,field10a,park11,sadeghi12a}:
\begin{eqnarray}
 L(\theta) &=& \sum_{n=1}^N \bigg( (x_n^T \theta) r_n - G(x_n^T\theta)
 \bigg)+ \mathrm{const(\theta)}\\
\label{EL}
	       &\approx& \left( \sum_{n=1}^N x_n^T r_n \right) \theta
- N\epec{G(x^T\theta)} \equiv \tilde{L}(\theta),
\end{eqnarray}
where the expectation is with respect to the distribution of $x$.  The
EL trades in the $O(KNp)$ cost of computing the nonlinear sum for the
cost of computing $\epec{G(x^T\theta)}$ at $K$ different values of
$\theta$, resulting in order $O(Kz)$ cost, where $z$ denotes the
cost of computing the expectation $\epec{G(x^T\theta)}$.  Thus the
nonlinear term of the EL can be be computed about $\frac{Np}{z}$ times
faster than the dominant term in the exact GLM log-likelihood.
Similar gains are available in computing the gradient and Hessian of
these terms with respect to $\theta$.

How hard is the integral $\epec{G(x^T\theta)}$ in practice?  I.e., how
large is $z$?  First, note that because $G$ only depends on the projection of
$x$ onto $\theta$, calculating this expectation only requires the
computation of a one-dimensional integral:
\begin{eqnarray}
\label{uniint}
\epec{G(x^T\theta)} &=& \int G(x^T\theta) p(x) dx =  \int G(q) \zeta_{\theta}(q) dq,
\end{eqnarray}
where $\zeta_{\theta}$ is the ($\theta$-dependent) distribution of the
one-dimensional variable $q = x^T\theta$.  If $\zeta_{\theta}$ is
available analytically, then we can simply apply standard
unidimensional numerical integration methods to evaluate the
expectation.

In certain cases this integral can be performed analytically.  Assume
(wlog) that $\epec{x} = 0$, for simplicity.  Consider the standard
regression case: recall that in this example 
\begin{eqnarray}
G(x^T\theta) &\propto& \frac{\theta^Txx^T \theta}{2},
\end{eqnarray}
implying that 
\begin{eqnarray}
\label{GaussExpec}
\epec{G(x^T\theta)} &=& \frac{\theta^TC\theta}{2},
\end{eqnarray}
where we have abbreviated $\epec{xx^T} = C$.  It should be noted that for this Gaussian example one only needs to compute the non-linear sum in the exact likelihood once, since $\sum_n G(x_n^T \theta) = \theta^T (\sum_n x_n x_n^T) \theta$ and $\sum_n x_n x_n^T$ can be precomputed.  However, as discussed in section \ref{sec:comp-effic-maxim}, if $C$ is chosen to have some special structure, e.g., banded, circulant, Toeplitz, etc., estimates of $\theta$ can still be computed orders of magnitude faster using the EL instead of the exact likelihood. 

The LNP model provides another example.  If $p(x)$ is Gaussian with
mean zero and covariance $C$, then 
\begin{eqnarray}
\label{eq:LNPapp}
\epec{G(x^T\theta)} &=& \int \exp(x^T\theta) \frac{1}{
  (2\pi)^\frac{p}{2} |C|^\frac{1}{2}} \exp \Big(-x^TC^{-1} x/2\Big) dx \\
				   &=&\exp\Big(\frac{\theta^TC \theta}{2}\Big),
\end{eqnarray}
where we have recognized the moment-generating function of the
multivariate Gaussian distribution.

Note that in each of the above cases, $\epec{G(x^T\theta)}$ depends
only on $\theta^TC \theta$.  This will always be the case (for any
nonlinearity $G(.)$) if $p(x)$ is elliptically symmetric, i.e.,
\begin{eqnarray}
	p(x) &=&  h(x^TC^{-1}x),
\end{eqnarray}
for some nonnegative function $h(.)$\footnote{Examples of such
  distributions include the multivariate normal and Student's-t, and
  exponential power families \cite{fang90a}.  Elliptically symmetric
  distributions are important in the theory of GLMs because they
  guarantee the consistency of the maximum likelihood estimator for
  $\theta$ even under certain cases of model misspecification; see
  \cite{paninski04b} for further discussion.}.  In this case we have
\begin{eqnarray}
\epec{G(x^T\theta)} &=& \int G(x^T\theta)h( x^TC^{-1}x) dx \\ &=& \int
G(y^T \theta') h(||y||^2_2) |C|^{1/2}dy, 
\end{eqnarray}
where we have made the change of variables $y=C^{-1/2}x,
\theta'=C^{1/2} \theta$.  Note that the last integral depends on
$\theta'$ only through its norm; the integral is invariant with
respect to transformations of the form $\theta' \rightarrow O
\theta'$, for any orthogonal matrix $O$ (as can be seen by the change
of variables $z=O^Ty$).  Thus we only need to compute this integral
once for all values of $||\theta'||^2_2 = \theta^T C \theta$, up to
some desired accuracy.  This can be precomputed off-line and stored in
a one-d lookup table before any EL computations are required, making
the amortized cost $z$ very small.

\begin{figure}[t!]
\begin{center}
\includegraphics[scale=0.9]{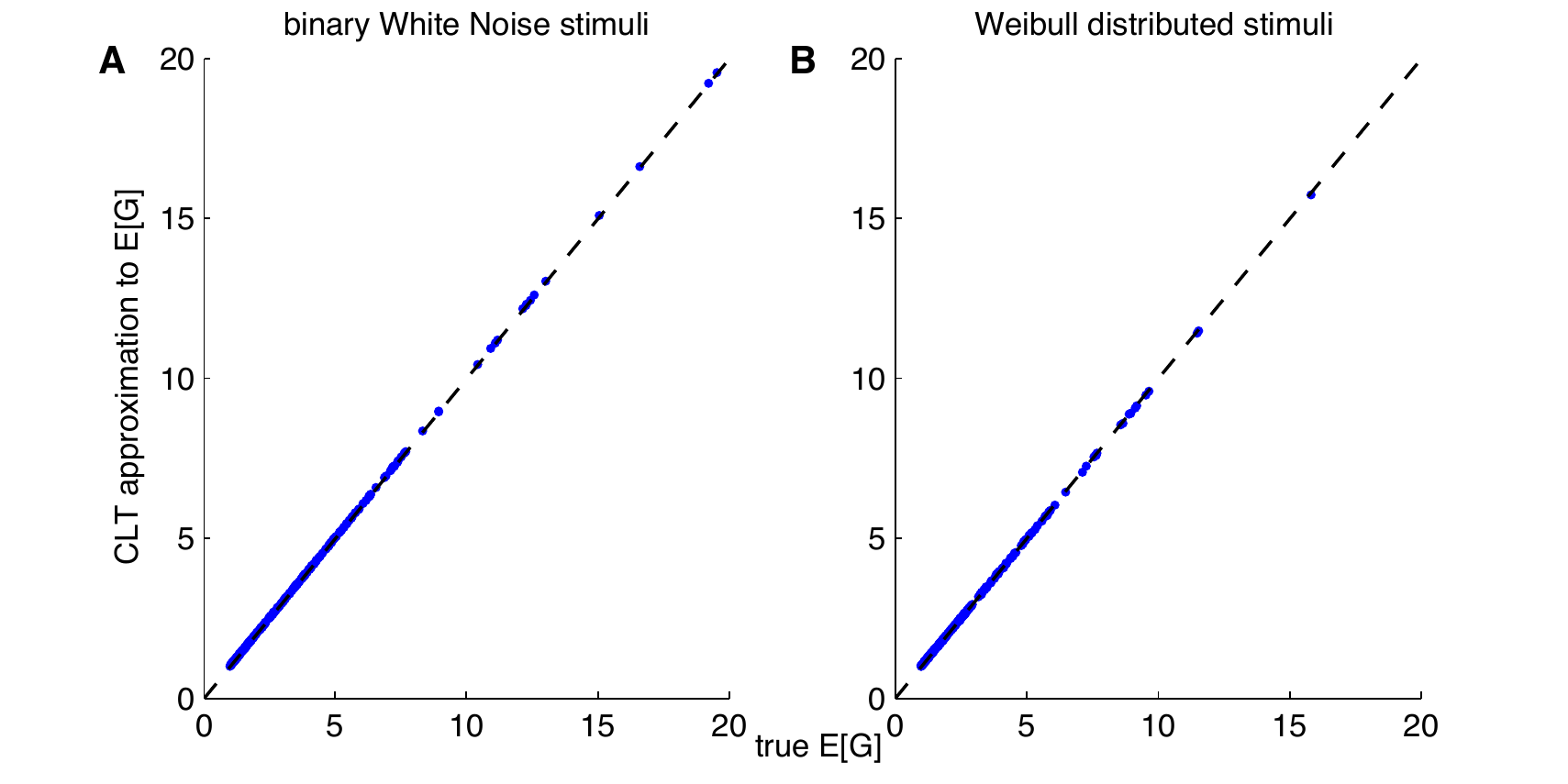}
\caption{The normal approximation for $\zeta_{\theta}$ is often quite
accurate for the computation of $\epec{G}$ (equation \ref{uniint} in
the text).  Vertical axis corresponds to central limit theorem
approximation of $\epec{G}$, and horizontal axis corresponds to the
true $\epec{G}$, computed numerically via brute-force Monte Carlo.  We
used a standard Poisson regression model here, corresponding to an
exponential $G$ function.  The stimulus vector $x$ is composed of
$600$ i.i.d. binary (A) or Weibull (B) variables; $x$ has mean $0.36$
in panel A, and the Weibull distribution in panel B has scale and
shape parameters $0.15$ and $0.5$, respectively.  Each dot corresponds
to a different value of the stimulus filter $\theta$.  These were zero-mean, Gaussian functions with randomly-chosen norm (uniformly distributed on the interval 0 to 0.5) and scale (found by taking the absolute value of a normally distributed variable with variance equal to 2).}
\label{cltacc}
\end{center}
\end{figure}

What if $p(x)$ is non-elliptical and we cannot compute
$\zeta_{\theta}$ easily?  We can still compute $\epec{G(x^T\theta)}$
approximately in most cases with an appeal to the central limit
theorem \cite{sadeghi12a}: we approximate $q = x^T \theta$ in equation
\ref{uniint} as Gaussian, with mean $\epec{\theta^Tx} =
\theta^T\epec{x}=0$ and variance $\mathrm{var}(\theta^Tx) =
\theta^TC\theta$.  This approximation can be justified by the classic
results of \cite{diaconis84a}, which imply that under certain
conditions, if $d$ is sufficiently large, then $\zeta_{\theta}$ is
approximately Gaussian for most projections $\theta$.  (Of course in
practice this approximation is most accurate when the vector $x$
consists of many weakly-dependent, light-tailed random variables and
$\theta$ has large support, so that $q$ is a weighted sum of many
weakly-dependent, light-tailed random variables.)  Thus, again, we can
precompute a lookup function for $\epec{G(x^T\theta)}$, this time over
the two-d table of all desired values of the mean and variance of $q$.
Numerically, we find that this approximation often works quite well;
Figure \ref{cltacc} illustrates the approximation for simulated
stimuli drawn from two non-elliptic distributions (binary white noise
stimuli in A and Weibull-distributed stimuli in B).

\subsection{Computational efficiency of maximum expected log-likelihood estimation for the LNP and Gaussian model}
\label{sec:comp-effic-maxim}

As a first application of the EL approximation, let us examine
estimators that maximize the likelihood or penalized likelihood.  We
begin with the standard maximum likelihood estimator,
\begin{eqnarray}
	\hat \theta_{MLE} &=& \arg \max_\theta \, \, L(\theta).
\end{eqnarray}
Given the discussion above, it is natural to maximize the expected
likelihood instead:
\begin{eqnarray}
\label{MargCost}
	\hat{\theta}_{MELE} &=& \arg \max_\theta \, \,   \tilde{L}(\theta),
\end{eqnarray} 
where ``MELE'' abbreviates ``maximum EL estimator.''  We expect that
the MELE should be computationally cheaper than the MLE by a factor of
approximately $Np/z$, since computing the EL is approximately a factor
of $Np/z$ faster than computing the exact likelihood.  In fact, in
many cases the MELE can be computed analytically while the MLE
requires a numerical optimization, making the MELE even faster.

Let's start by looking at the standard regression (Gaussian noise)
case.  The EL here is proportional to
\begin{eqnarray}
\label{gaussmarg}
\tilde{L}(\theta) &\propto& \theta^TX^Tr - N\frac{\theta^TC
\theta}{2},
\end{eqnarray}
where we have used equation \ref{GaussExpec} and defined $X = (x_1,
..., x_N)^T$.  This is a quadratic function of $\theta$; optimizing
directly, we find that the MELE is given by
\begin{eqnarray}
\label{MEG}
	\margNR = (NC)^{-1} X^Tr.
\end{eqnarray}
Meanwhile, the MLE here is of the standard least-squares form
\begin{eqnarray}
\mle=(X^TX)^{-1}X^Tr,
\end{eqnarray}
assuming $X^TX$ is invertible (the solution is non-unique otherwise).

The computational cost of determining both estimators is determined by
the cost of solving a $p$-dimensional linear system of equations; in
general, this requires $O(p^3)$ time.  However, if $C$ has some
special structure, e.g., banded, circulant, Toeplitz, etc.\ (as is
often the case in real neuroscience experiments), this cost can be
reduced to $O(p)$ (in the banded case) or $O(p\log(p))$ (in the
circulant case) \cite{golub96a}.  The MLE will typically not enjoy
this decrease in computational cost, since in general $X^TX$ will be
unstructured even when $C$ is highly structured.  (Though
counterexamples do exist; for example, if $X$ is highly sparse, then
$X^TX$ may be sparse even if $C$ is not, for sufficiently small $N$.)

As another example, consider the LNP model.  Somewhat surprisingly,
the MELE can be computed analytically for this model \cite{park11} if
$p(x)$ is Gaussian and we modify the model slightly to include an
offset term so that the Poisson rate in the $n$-th time bin is given
by
\begin{eqnarray}
  \lambda_n = \exp(\theta_0 + x_n ^T \theta),
\end{eqnarray}
with the likelihood and EL modified appropriately.  The details are
provided in \cite{park11} and also, for completeness, the methods section of this paper; the key result is that if one first optimizes the EL
(equation \ref{EL}) with respect to the offset $\theta_0$ and then
substitutes the optimal $\theta_0$ back into the EL, the resulting
``profile'' expected log-likelihood $\max_{\theta_0}
\tilde{L}(\theta,\theta_0)$ is a quadratic function of $\theta$, which
can be optimized easily to obtain the MELE:
\begin{eqnarray}
\label{rsta}
\margNR &=& \arg \max_{\theta} \, \, \max_{\theta_0}
\tilde{L}(\theta,\theta_0) \\ &=& \arg \max_{\theta} \, \,
\theta^TX^Tr -\sum_{n=1}^Nr_n\frac{\theta^TC\theta}{2} \\ &=& \left(
(\sum_n r_n)C \right)^{-1} X^Tr.
\end{eqnarray}
Note that this is essentially the same quadratic problem as in the
Gaussian case (equation \ref{gaussmarg}) with the total number of
spikes $\sum_n r_n$ replacing the number of samples $N$ in equation
\ref{gaussmarg}.  In the neuroscience literature, the function
$\frac{X^Tr}{\sum_n r_n}$ is referred to as the spike-triggered
average, since if time is discretized finely enough so that the
entries of $r$ are either $0$ or $1$, the product $X^T\frac{r}{\sum_n r_n}$ is simply an
average of the stimulus conditioned on the occurrence of a `spike'
($r=1$).  The computational cost for computing $\margNR$ here is
clearly identical to that in the Gaussian model (only a simple linear
equation solve is required), while to compute the MLE we need to
resort to numerical optimization methods, costing $O(KNp)$, with $K$
typically depending superlinearly on $p$.  The MELE can therefore be
orders of magnitude faster than the MLE here if $Np$ is large,
particularly if $C$ has some structure that can be exploited.  See
\cite{park11,sadeghi12a} for further discussion.

What about estimators that maximize a penalized likelihood?  Define
the maximum penalized expected log-likelihood estimator (MPELE)
\begin{eqnarray}
\label{margMAP}
	\marg &=& \arg \max_{\theta} \, \,  \tilde{L}(\theta) + \log(f(\theta)),
\end{eqnarray}
where $\log(f(\theta))$ represents a penalty on $\theta$; in many
cases $f(\theta)$ has a natural interpretation as the prior
distribution of $\theta$.  We can exploit special structure in $C$
when solving for the MPELE as well.  For example, if we use a
mean-zero (potentially improper) Gaussian prior, so that
$\log(f(\theta)) = - \frac{1}{2} \theta ^T R \theta$ for some positive
semidefinite matrix $R$, the MPELE for the LNP model is again a
regularized spike-triggered average (see \cite{park11} and methods)
\begin{eqnarray}
\label{regrsta}
 \marg &=&  \Big(C + \frac{R}{\sum_n r_n}\Big)^{-1}\frac{X^Tr }{ \sum_n r_n}.
\end{eqnarray}
For general matrices $R$ and $C$, the dominant cost of computing
$\marg$ will be $O(Np + p^3)$.  (The exact maximum a posteriori (MAP)
estimator has cost comparable to the MLE here, $O(KNp)$.)  Again, when
$C$ and $R$ share some special structure, e.g. $C$ and $R$ are both
circulant or banded, the cost of $\marg$ drops further.  

If we use a sparsening L$_1$ penalty instead
\cite{David07,calabrese11a}, i.e., $\log(f(\theta)) = - \lambda \|
\theta\|_1$, with $\lambda$ a scalar, $\marg$ under a Gaussian model
is defined as
\begin{eqnarray}
\label{gaussL1}
\marg &=& \arg \max_{\theta} \, \, \theta^TX^Tr - N\frac{\theta^TC
  \theta}{2} - \lambda \| \theta\|_1;
\end{eqnarray}
the MPELE under an LNP model is of nearly identical form.  If $C$ is a
diagonal matrix, classic results from subdifferential calculus
\cite{nesterov04a} show that $\marg$ is a solution to equation
\ref{gaussL1} if and only if $\marg$ satisfies the subgradient
optimality conditions
\begin{eqnarray}
  -NC_{jj} (\marg)_j + (X^Tr)_j  &=& \lambda ~ \mathrm{sign}(\marg)_j \, \, \, \mathrm{if} \, \, (\marg)_j \neq 0 \\
  \Big| -NC_{jj} (\marg)_j + (X^Tr)_j \Big| &\leq& \lambda \, \, \, \mathrm{otherwise},
\end{eqnarray}
for $j =1, ..., p$.  The above equations imply that $\marg$ is a
soft-thresholded function of $X^Tr$: $(\marg)_j = 0$ if  $|(X^Tr)_j|
\leq \lambda$, and otherwise
\begin{eqnarray}
\label{solL1marg}
(\marg)_j &=& \frac{1}{NC_{jj}} \Big((X^Tr)_j -\lambda ~ \mathrm{sign}(\marg)_j\Big),
\end{eqnarray}
for $j=1, ...,p$.  Note that equation \ref{solL1marg} implies that we
can independently solve for each element of $\marg$ along all values
of $\lambda$ (the so-called regularization path).  Since only a single
matrix-vector multiply ($X^Tr$) is required, the total complexity in
this case is just $O(Np)$.  Once again, because $X^TX$ is typically
unstructured, computation of the exact MAP is generally much more
expensive.

When $C$ is not diagonal we can typically no longer solve equation
\ref{gaussL1} analytically.  However, we can still solve this equation
numerically, e.g., using interior-point methods \cite{CONV04}.
Briefly, these methods solve a sequence of auxiliary, convex problems
whose solutions converge to the desired vector.  Unlike problems with
an L$_1$ penalty, these auxiliary problems are constructed to be
smooth, and can therefore be solved in a small number of iterations
using standard methods (e.g., Newton-Raphson (NR) or conjugate
gradient (CG)).  Computing the Newton direction requires a linear
matrix solve of the form $(C+D)\theta=b$, where $D$ is a diagonal
matrix and $b$ is a vector.  Again, structure in $C$ can often be
exploited here; for example, if $C$ is banded, or diagonal plus
low-rank, each Newton step requires $O(p)$ time, leading to
significant computational gains over the exact MAP.

To summarize, because the population covariance $C$ is typically more
structured than the sample covariance $X^TX$, the MELE and MPELE can
often be computed much more quickly than the MLE or the MAP estimator.
We have examined penalized estimators based on L$_2$ and L$_1$
penalization as illustrative examples here; similar conclusions hold
for more exotic examples, including group penalties, rank-penalizing
penalties, and various combinations of L$_2$ and L$_1$ penalties.

\subsection{Analytic comparison of the accuracy of EL estimators with the accuracy of maximum-likelihood estimators}
\label{mathanaly}

We have seen that EL-based estimators can be fast.  How accurate are
they, compared to the corresponding MLE or MAP estimators?  First,
note that the MELE inherits the classical consistency properties of
the MLE; i.e., if the model parameters are held fixed and the amount
of data $N$ tends to infinity, then both of these estimators will
recover the correct parameter $\theta$, under suitable conditions.
This result follows from the classical proof of the consistency of the
MLE \cite{vaart98a} if we note that both $(1/N) L(\theta)$ and $(1/N)
\tilde{L} (\theta)$ converge to the same limiting function of $\theta$.

To obtain a more detailed view, it is useful to take a close look at
the linear regression model, where we can analytically calculate the
mean-squared error (MSE) of these estimators.  Recall that we assume
\begin{eqnarray}
\label{gmodel1}
  r|x &\sim& \mathcal{N}(x^T\theta,I),\\ 
  \label{gmodel2}
 x &\sim& \mathcal{N}(0,I).
\end{eqnarray}
(For convenience we have set $\sigma^2=1$.)  We derive the following
MSE formulas in the methods:
\begin{eqnarray}
        \label{mmsea}
\epec{ \| \margNR- \theta \|^2_2}           &=&  \frac{ \theta^T \theta + p( \theta^T \theta + 1) }{N}, \\
\epec{  \| \mle- \theta \|^2_2}  &=& \frac{p}{N-p-1};
\label{mmse_MLa}
\end{eqnarray}
see \cite{shaffer91a} for some related results.  In the classical
limit, for which $p$ is fixed and $N \to \infty$ (and the MSE of both
estimators approaches zero), we see that, unless $\theta = 0$, the MLE outperforms the MELE:
\begin{eqnarray}
  \lim_{N \to \infty} N \epec{ \| \margNR- \theta \|^2_2} > \lim_{N
  \to \infty} N \epec{ \| \mle- \theta \|^2_2} = p.
\end{eqnarray}

However, for many applications (particularly the neuroscience
applications we will focus on below), it is more appropriate to
consider the limit where the number of samples and parameters are
large, both $N \rightarrow \infty$ and $p \rightarrow \infty$, but
their ratio $\frac{p}{N} = \rho$ is bounded away from zero and
infinity.  In this limit we see that
\begin{eqnarray}
        \label{mmse}
\epec{ \| \margNR- \theta \|^2_2}           &\rightarrow& \rho( \theta^T \theta + 1) \\
\label{mmse_ML}
\epec{  \| \mle- \theta \|^2_2}  &\rightarrow& \frac{\rho}{1-\rho}.
 \end{eqnarray}
See figure \ref{finitesamp} for an illustration of the accuracy of
this approximation for finite $N$ and $p$.

\begin{figure}[t!]
\begin{center}
\includegraphics[scale = 0.9]{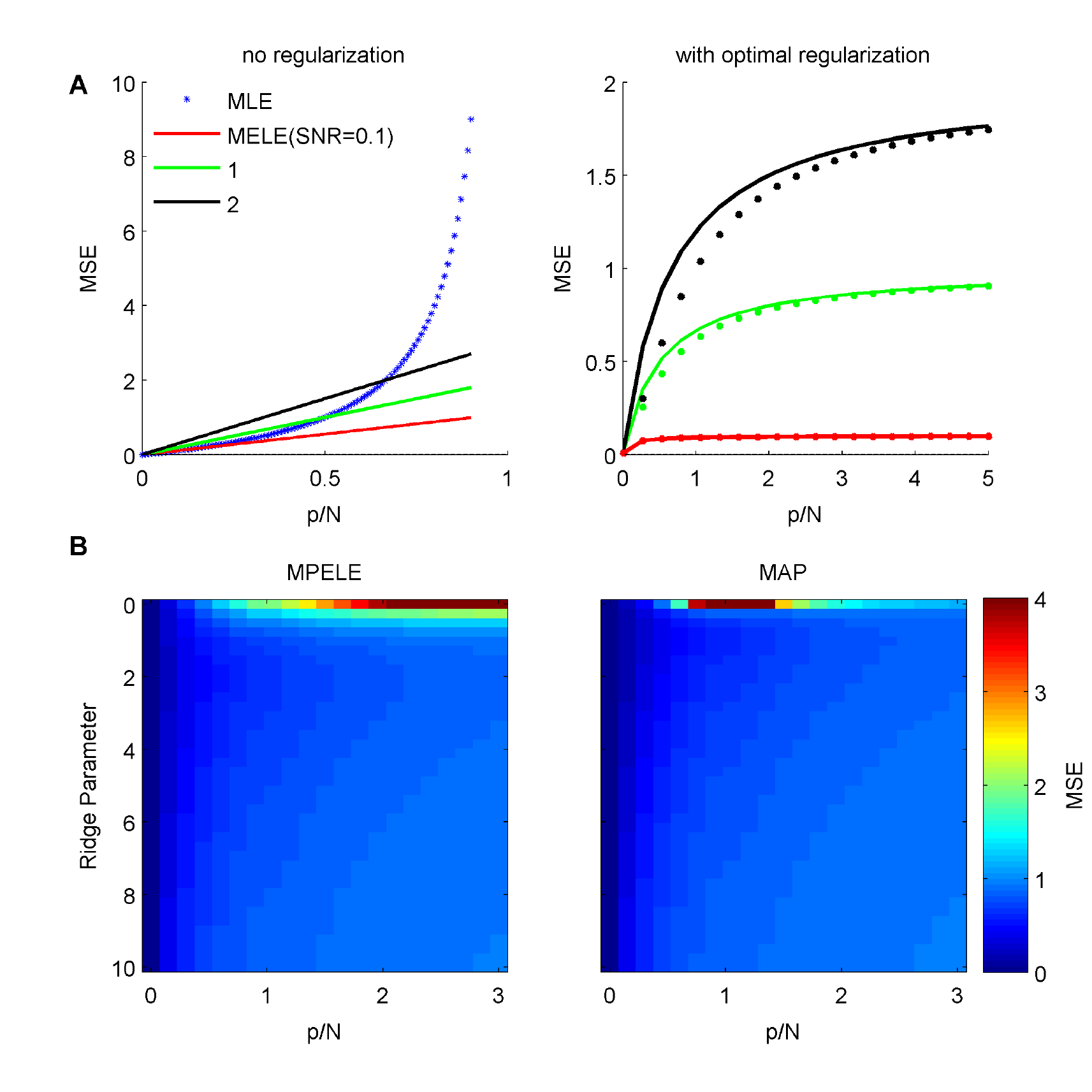}
\caption{Comparing the accuracy of the MAP and MPELE in the standard
  linear regression model with Gaussian noise.  A.) (left) The mean
  squared error (MSE) for the MELE (solid lines) and the MLE (dotted
  line) is shown as a function of $p/N$ the ratio of the number of
  parameters to number of samples.  We plot results for the asymptotic
  case where the number of samples and dimensions goes to infinity but
  the ratio $p/N$ remains finite.  Different colors denote different
  values for the true filter norm; recall that the MSE of the MLE is
  independent of the true value of $\theta$, since the MLE is an
  unbiased estimator of $\theta$ in this model.  The MLE mean squared
  error is larger than that of the MELE when $p/N$ is large.  B.) MSE
  for both estimators when L2 regularization is added.  The MSE is
  similar for both estimators for a large range of ridge parameters
  and values of $p/N$. A.)  (right) For each value of $p/N$, separate
  ridge parameters are chosen for the MPELE and MAP estimators to
  minimize their respective mean squared errors.  Solid curves
  correspond to MPELE (as in left panel); dotted curves to MAP
  estimates.  The difference in performance between the two
  optimally-regularized estimators remains small for a wide range of
  values of SNR and $p/N$.  Similar results are observed numerically
  in the Poisson regression (LNP) case (data not shown).}
\label{toymse}
\end{center}
\end{figure}

Figure \ref{toymse}A (left panel) plots these limiting MSE curves as a
function of $\rho$.  Note that we do not plot the MSE for values of
$\rho > 1$ because the MLE is non-unique when $p$ is greater than $N$;
also note that eq.~(\ref{mmse_ML}) diverges as $\rho \nearrow 1$,
though the MELE MSE remains finite in this regime.  We examine these
curves for a few different values of $\theta^T\theta$; note that since
$\sigma^2=1$, $\theta^T\theta$ can be interpreted as the signal
variance divided by the noise variance, i.e., the signal-to-noise
ratio (SNR):
\begin{eqnarray}
	SNR &=& \frac{\epec{\theta^Tx^Tx\theta}}{\sigma^2}\\
		&=& \frac{\theta^T\theta}{1}.
\end{eqnarray}
The second line follows from the fact that we choose stimuli with
identity covariance.  The key conclusion is that the MELE outperforms
the MLE for all $\rho > \frac{SNR}{1+SNR}$.  (This may seem
surprising, since for a given $X$, a classic result in linear
regression is that the MLE has the lowest MSE amongst all unbiased
estimators of $\theta$ \cite{bickel07a}.  However, the MELE is biased
given $X$, and can therefore have a lower MSE than the MLE by having a
smaller variance.)

What if we examine the penalized versions of these estimators?  In the
methods we calculate the MSE of the MAP and MPELE given a simple ridge
penalty of the form $\log(f(\theta)) = - \frac{R}{2} \| \theta\|_2^2$,
for scalar $R$.  Figure \ref{toymse}B (top panel) plots the MSE for
both estimators (see equations \ref{margmap-asy}, \ref{map-asy} in the
methods for the equations being plotted) as a function of $R$ and
$\rho$ for an SNR value of 1.  Note that we now plot MSE values for
$\rho > 1$ since regularization makes the MAP solution unique.  We see
that the two estimators have similar accuracy over a large region of
parameter space.  For each value of $\rho$ we also compute each
estimator's optimal MSE --- i.e., the MSE corresponding to the the
value of $R$ that minimizes each estimator's MSE.  This is plotted in
Figure \ref{toymse}A (right panel).  Again, the two estimators perform
similarly.

In conclusion, in the limit of large but comparable $N$ and $p$, the MELE
outperforms the MLE in low-SNR or high-$(p/N)$ regimes.  The ridge-regularized estimates (the MPELE and MAP) perform similarly across a broad range of $(p/N)$ and regularization values (Figure \ref{toymse}B).  These analytic results motivate the applications (using non-Gaussian GLMs) on real data treated in the next section.  

\subsection{Fast methods for refining maximum expected log-likelhood estimators to obtain MAP accuracy}
\label{secM1}

In settings where the MAP provides a more accurate estimator than the
MPELE, a reasonable approach is to use $\marg$ as a quickly-computable
initializer for optimization algorithms used to compute $\map$.  An
even faster approach would be to initialize our search at $\marg$,
then take just enough steps towards $\map$ to achieve an estimation
accuracy which is indistinguishable from that of the exact MAP
estimator.  (Related ideas based on stochastic gradient ascent methods
have seen increasing attention in the machine learning literature
\cite{bottou98a,Welling11}.) We tested this idea on real data, by
fitting an LNP model to a population of ON and OFF parasol ganglion
cells (RGCs) recorded \textit{in vitro}, using methods similar to
those described in \cite{shlens06a,pillow08a}; see these earlier
papers for full experimental details.  The observed cells responded to
either binary white-noise stimuli or naturalistic stimuli
(spatiotemporally correlated Gaussian noise with spatial
correlations having a 1/F power spectrum and temporal correlations
defined by a first-order autoregressive process; see methods for
details).  As described in the methods, each receptive field was
specified by $810$ parameters, with $\frac{p}{N}= 0.021$.  For the
MAP, we use a simple ridge penalty of the form $\log(f(\theta)) = -
\frac{R}{2} \| \theta\|_2^2$.

Many iterative algorithms are available for ascending the posterior to
approximate the MAP.  Preconditioned conjugate gradient ascent (PCG)
\cite{Shewchuk94} is particularly attractive here, for two reasons.
First, each iteration is fairly fast: the gradient requires $O(Np)$
time, and multiplication by the preconditioner turns out to be fast,
as discussed below.  Second, only a few iterations are necessary,
because the MPELE is typically fairly close to the exact MAP in this
setting (recall that $\marg \to \map$ as $N/p \to \infty$), and we
have access to a good preconditioner, ensuring that the PCG iterates
converge quickly.  We chose the inverse Hessian of the EL evaluated at
the MELE or MPELE as a pre-conditioner.  In this case, using the same
notation as in equations \ref{rsta} and \ref{regrsta}, the
preconditioner is simply given by $(C\sum_n r_n)^{-1}$ or $(C\sum_n
r_n+RI)^{-1}$.  Since the EL Hessian provides a good approximation for
the log-likelihood Hessian, the preconditioner is quite accurate; since
the stimulus covariance $C$ is either proportional to the identity (in
the white-noise case) or of block-Toeplitz form (in the
spatiotemporally-correlated case), computation with the preconditioner
is fast ($O(p)$ or $O(p \log p)$, respectively).

\begin{figure}[t!]
\begin{center}
\includegraphics[scale=0.99]{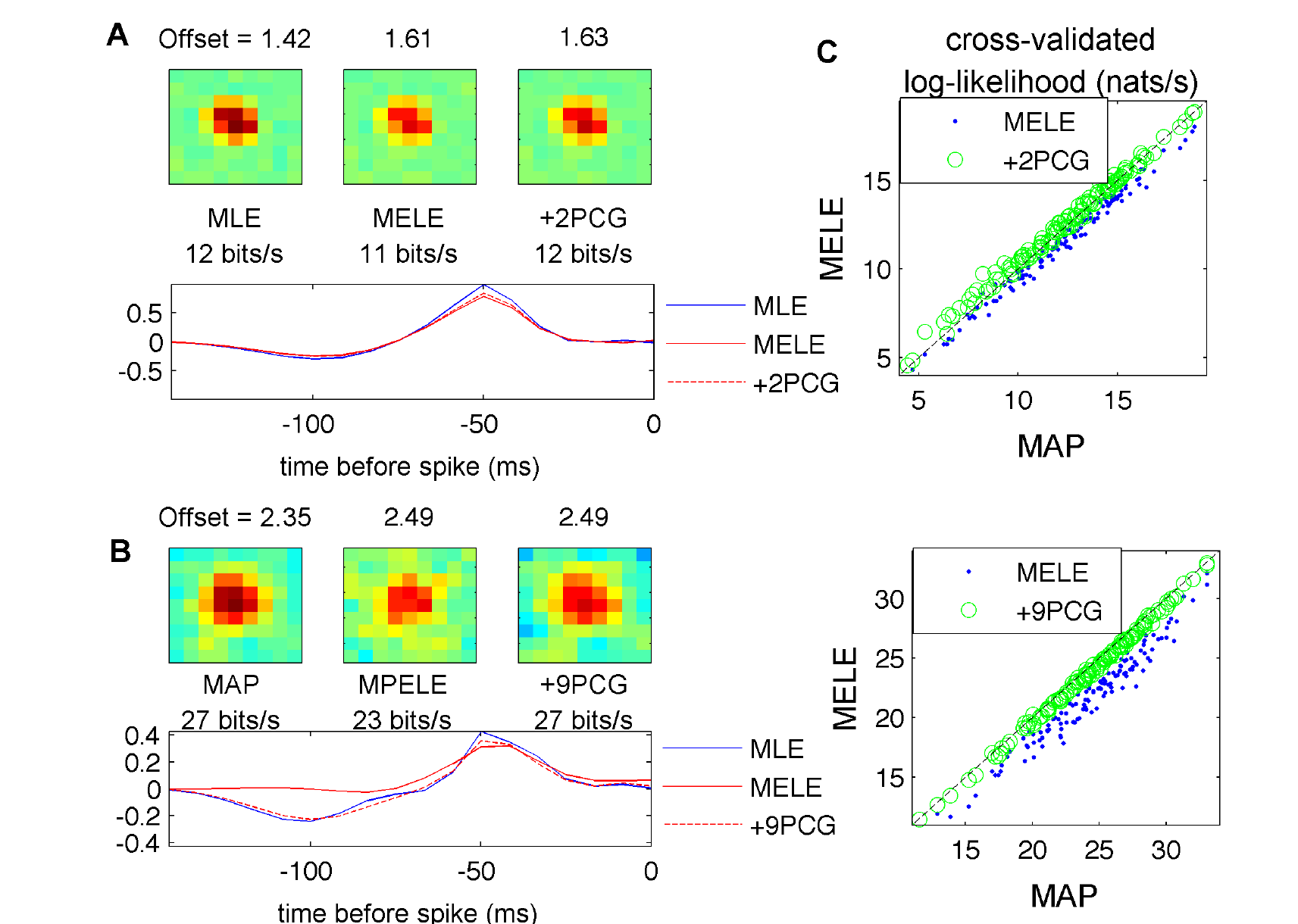}
\caption{The MPELE provides a good initializer for finding approximate
MAP estimators using fast gradient-based optimization methods in the
LNP model.  A.)  The spatiotemporal receptive field of a typical
retinal ganglion cell (RGC) in our database responding to binary
white-noise stimuli and fit with a linear-nonlinear Poisson (LNP)
model with exponential non-linearity, via the MLE.  The receptive
field of the same cell fit using the MELE is also plotted.  The
goodness-of-fit of the MLE (measured in terms of cross-validated
log-likelihood; see methods) is slightly higher than that of the MELE
(12 versus 11 bits/s).  However, this difference disappears after a
couple pre-conditioned conjugate gradient (PCG) iterations using the
true likelihood initialized at the MELE (see label +2PCG).  Note that
we are only showing representative spatial and temporal slices of the
9x9x10 dimensional receptive field, for clarity.  B.) Similar results
hold when the same cell responds to 1/f correlated Gaussian noise
stimuli (for correlated Gaussian responses, the MAP and MPELE are both
fit with a ridge penalty).  In this case 9 PCG iterations sufficed to
compute an estimator with a goodness-of-fit equal to that of the MAP.
C.)  These results are consistent over the observed population. (top)
Scatterplot of cross-validated log-likelihood across a population of 91 cells,
responding to binary white-noise stimuli, each fit independently using
the MLE, MELE or a couple PCG iterations using the true likelihood and
initialized at the MELE.  (bottom) Log-likelihood for the same
population, responding to 1/f Gaussian noise, fit independently using
an L2 regularized MAP, MPELE or 9 PCG iterations using the true
regularized likelihood initialized at the MPELE.}
\label{LNPdata}
\end{center}
\end{figure}

For binary white-noise stimuli we find that the MELE (given by
equation \ref{rsta} with $C=\mathbf{I}$) and MLE yield similar filters
and accuracy, with the MLE slightly outperforming the MELE (see figure
\ref{LNPdata}A).  Note that in this case, $\margNR$ can be computed
quickly, $O(pN)$, since we only need to compute the matrix-vector
multiplication $X^Tr$.  On average across a population of 126 cells,
we find that terminating the PCG algorithm, initialized at the MELE,
after just two iterations yielded an estimator with the same accuracy
as the MAP.  To measure accuracy we use the cross-validated
log-likelihood (see methods).  It took about $15 \times$ longer to
compute the MLE to default precision than the PCG-based approximate
MLE (88$\pm$2 vs 6$\pm$0.1 seconds on an Intel Core 2.8 GHz processor
running Matlab; all timings quoted in this paper use the same
computer).  In the case of spatiotemporally correlated stimuli (with
$\marg$ given by equation \ref{regrsta}), we find that 9 PCG
iterations are required to reach MAP accuracy (see figure
\ref{LNPdata}B); the MAP estimator was still slower to compute by a
significant factor (107$\pm$8 vs 33$\pm$1 seconds).

\subsubsection{Scalable modeling of interneuronal dependencies}

So far we have only discussed models which assume conditional
independence of responses given an external stimulus.  However, it is
known the predictive performance of the GLM can be improved in a
variety of brain areas by allowing some conditional dependence between
neurons (e.g., \cite{truccolo05a,pillow08a}; see \cite{vidne11a} for a
recent review of related approaches).  One convenient way to
incorporate these dependencies is to include additional covariates in
the GLM.  Specifically, assume we have recordings from $M$ neurons and
let $r_i$ be the vector of responses across time of the $i$th neuron.
Each neuron is modeled with a GLM where the weighted covariates,
$x_n^T\theta_i$, can be broken into an offset term, an external
stimulus component $x^s$, and a spike history-dependent component
\begin{eqnarray}
\label{hdGLM}
    x_n^T\theta_i &=& \theta_{i0} + (x^s)^T\theta^s_i + \sum_{j=1}^M
    \sum_{k=1}^\tau r_{j,n-k}\theta_{ijk}^H,
\end{eqnarray}
for $n=1,...,N$, where $\tau$ denotes the maximal number of lags used
to predict the firing rate of neuron $i$ given the activity of the the
observed neurons indexed by $j$.  Note that this is the same model as
before when $\theta_{ijk}^H = 0$ for $j=1,2,...., M$; in this special
case, we recover an inhomogeneous Poisson model, but in general, the
outputs of the coupled GLM will be non-Poisson.  A key challenge for
this model is developing estimation methods that scale well with the
number of observed neurons; this is critical since, as discussed
above, experimental methods continue to improve, resulting in rapid
growth of the number of simultaneously observable neurons during a
given experiment \cite{stevenson11a}.

One such scalable approach uses the MELE of an LNP model to fit a
fully-coupled GLM with history terms.  The idea is that estimates of
$\theta^s$ fit with $\theta_{ijk}^H = 0$ will often be similar to
estimates of $\theta^s$ without this hard constraint.  Thus we can
again use the MELE (or MPELE) as an initializer for $\theta^s$, and
then use fast iterative methods to refine our estimate, if necessary.
More precisely, to infer $\theta_i$ for $i=1,2,...., M$, we first
estimate $\theta_i^s$, assuming no coupling or self-history terms
($\theta_{ijk}^H = 0$ for $j=1,2,...., M$) using the MELE (equation
\ref{rsta} or a regularized version).  We then update the history
components and offset by optimizing the GLM likelihood with the
stimulus filter held fixed up to a gain factor, $\alpha_i$, and the
interneuronal terms $\theta^H_{ijk}=0$ for $i \neq j$:
\begin{eqnarray}
\label{cpfi}
(\theta_{i0},\alpha_i,\theta_i^H) &=&
\arg\max_{\substack{(\theta_{0},\alpha,\theta^H_{ijk} = 0 ~ \forall i
    \neq j)}} \, \,
L\Big((\theta_0, \alpha \marg, \theta^H) \Big) + \log\Big(f(\theta_0,
\alpha \marg, \theta^H) \Big).
\end{eqnarray}
Holding the shape of the stimulus filter fixed greatly reduces the
number of free parameters, and hence the computational cost, of
finding the history components.  Note that all steps so far scale
linearly in the number of observed neurons $M$.  Finally, perform a
few iterative ascent steps on the full posterior, incorporating a
sparse prior on the interneuronal terms (exploiting the fact that
neural populations can often be modeled as weakly conditionally
dependent given the stimulus; see \cite{pillow08a} and \cite{yuri2011}
for further discussion).  If such a sparse solution is available, this
step once again often requires just $O(M)$ time, if we exploit
efficient methods for optimizing posteriors based on
sparsity-promoting priors \cite{friedman10a,zhang11a}.

We investigated this approach by fitting the GLM specified by equation
\ref{hdGLM} to a population of 101 RGC cells responding to binary
white-noise using 250 stimulus parameters and 105 parameters related
to neuronal history (see methods for full details;
$\frac{p}{N}=0.01$).  We regularize the coupling history components
using a sparsity-promoting L$_1$ penalty of the form $\lambda \sum_j
|\theta^H_{ij}|$; for simplicity, we parametrize each coupling filter
$\theta^H_{ij}$ with a single basis function, so that it is not
necessary to sum over many $k$ indices.  (However, note that
group-sparsity-promoting approaches are also straightforward here
\cite{pillow08a}.)  We compute our estimates over a large range of the
sparsity parameter $\lambda$ using the ``glmnet'' coordinate ascent
algorithm discussed by \cite{friedman10a}.

\begin{figure}[t!]
\begin{center}
\includegraphics[scale=0.8]{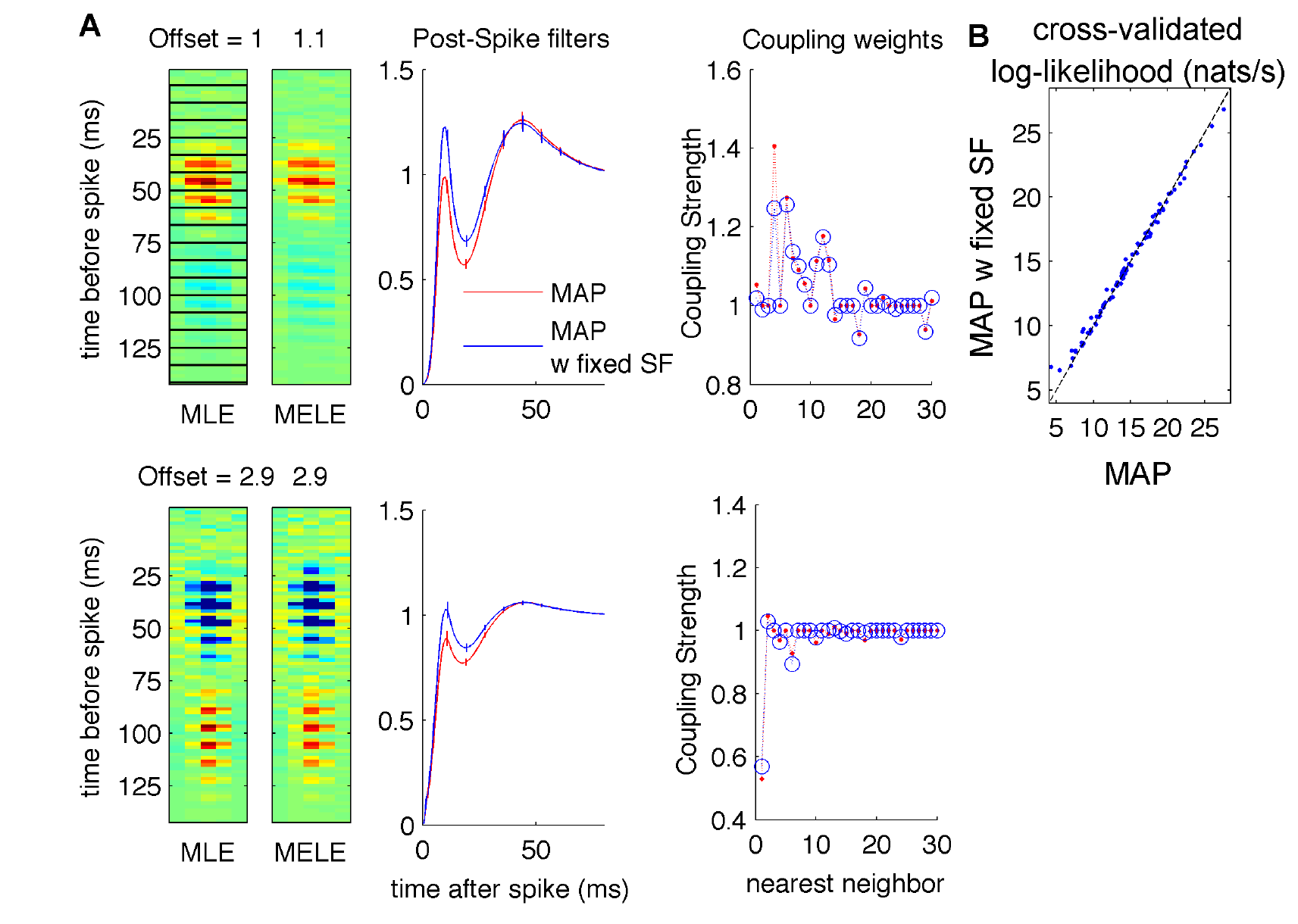}
\caption{Initialization with the MPELE using an LNP model, then
  coordinate descent using a sparsity-promoting prior, efficiently
  estimates a full, coupled, non-Poisson neural population GLM.  A.)
  Example stimulus, self-history, and coupling filters for two
  different RGC cells (top and bottom rows).  The stimulus filters are
  laid out in a rasterized fashion to map the three-dimensional
  filters (two spatial dimensions and one temporal dimension) onto the
  two-dimensional representation shown here.  Each filter is shown as a stack (corresponding to the 
  temporal dimension) of two dimensional spatial filters, which we outline in black in the top left 
  to aid the visualization.  MAP parameters are found
  using coordinate descent to optimize the exact GLM log-likelihood
  with L$_1$-penalized coupling filters (labeled MAP).  Fast estimates
  of the self-history (see methods for details of errorbar computations) and coupling filters are found by running the
  same coordinate descent algorithm with the stimulus filter (SF)
  fixed, up to a gain, to the MELE of an LNP model (labeled `w fixed
  SF'; see text for details).  Note that estimates obtained using
  these two approaches are qualitatively similar.  In particular, note
  that both estimates find coupling terms that are largest for
  neighboring cells, as in \cite{pillow08a,vidne11a}.  We do not plot the 
  coupling weights for cells 31-100 since most of these are zero or small.  B.) Scatterplot
  comparing the cross-validated log-likelihood over 101 different RGC
  cells show that the two approaches lead to comparable predictive
  accuracy.}
\label{glmnet}
\end{center}
\end{figure}

Figure \ref{glmnet}A compares filter estimates for two example cells
using the fast approximate method and the MAP.  The filter estimates
are similar (though not identical); both methods find the same
coupled, nearest neighbor cells.  Both methods achieve the same
cross-validated prediction accuracy (Fig.~\ref{glmnet}B).  The fast
approximate method took an average of 1 minute to find the entire
regularization path; computing the full MAP path took 16 minutes on
average.

\subsection{Marginal likelihood calculations}

In the previous examples we have focused on applications that require
us to maximize the (penalized) EL.  In many applications we are more
interested in integrating over the likelihood, or sampling from a
posterior, rather than simply optimization.  The EL approximation can
play an important role in these applications as well.  For example, in
Bayesian applications one is often interested in computing the
marginal likelihood: the probability of the data, $p(r)$, with the
dependence on the parameter $\theta$ integrated out
\cite{gelman03a,KR95}.  This is done by assigning a prior
distribution $f(\theta|R)$, with its own ``hyper-parameters'' $R$, over
$\theta$ and calculating
 \begin{eqnarray}
  \label{evidence}
	 F(R) \equiv p(r|x_1, ..., x_N,R) &=& \int p(r,\theta|x_1, ..., x_N,R) d\theta
	                                     = \int p(r|\theta,x_1,
                                             ..., x_N) f(\theta |R) d\theta.
	\end{eqnarray}
Hierarchical models in which $R$ sets the prior over $\theta_i$ for
many neurons $i$ simultaneously are also useful \cite{BKW05}; the
methods discussed below extend easily to this setting.

We first consider a case for which this integral is analytically
tractable.  We let the prior on $\theta$ be Gaussian, and use the
standard regression model for $r|\theta$:
\begin{eqnarray}
p(r,\theta|X,R) &=& p(r|X,\theta)f(\theta|R)\\
			&=& \mathcal{N}(X\theta,\sigma^2\mathbf{I}) \mathcal{N}(0,R^{-1}),
\end{eqnarray}
where $\mathbf{I}$ is the identity matrix.  Computing the resulting
Gaussian integral, we have
\begin{eqnarray}
\label{intevidence}
\log F(R) &=&  \frac{1}{2} \log \Big( \det(\Sigma R) \Big) + \frac{r^TX\Sigma X^Tr}{2\sigma^4} + \mathrm{const}(R),\\
 \Sigma &=& (X^TX\sigma^2 + R)^{-1}.
\end{eqnarray}
If $R$ and $X^TX$ do not share any special structure, each evaluation
of $F(R)$ requires $O(p^3)$ time.

On the other hand, if we approximate $p(r|X,\theta)$ by the the
expected likelihood, the integral is still tractable and has the same
form as equation \ref{intevidence}, but with $\Sigma = (NC\sigma^2
+ R)^{-1}$ assuming $\mathrm{Cov}[x]=C$.  In many cases the resulting
$F(R)$ calculations are faster: for example, if $C$ and $R$ can be
diagonalized (or made banded) in a convenient shared basis, then
$F(R)$ can often be computed in just $O(p)$ time.

When the likelihood or prior is not Gaussian we can approximate the
marginal likelihood using the Laplace approximation \cite{KR95}
\begin{eqnarray}
\label{lapp}
\log F(R)  &\approx& \log p(\theta_{MAP}|R) - \frac{1}{2} \log(\det(-H(\theta_{MAP}))) \\
\nonumber
		   &=&   \theta_{MAP}^T X^Tr - \sum_n G(x_n^T\theta_{MAP}) + \log\Big( f(\theta_{MAP}|R)\Big) - \frac{1}{2} \log(\det(-H(\theta_{MAP}))),		   
\end{eqnarray}
again neglecting factors that are constant with respect to $R$.
$H(\theta_{MAP})$ is the posterior Hessian
 \begin{eqnarray}
\label{eq:marg-like-hess}
H_{ij} &=& \frac{\partial^2}{\partial \theta_i \partial \theta_j} \Big(- \sum_n G(x_n^T\theta)  + \log\Big( f(\theta|R)\Big) \Big),
\end{eqnarray}
evaluated at $\theta = \theta_{MAP}$.  We note that there are other
methods for approximating the integral in equation \ref{evidence},
such as Evidence Propagation \cite{MINKAPHD,bishop06a}; we leave an
exploration of EL-based approximations of these alternative methods
for future work.

We can clearly apply the EL approximation in
eqs.~(\ref{lapp}-\ref{eq:marg-like-hess}).  For example, consider the
LNP model, where we can approximate $\epec{G(\theta_0+x^T\theta)}$ as
in equation \ref{eq:LNPapp}; if we use a normal prior of the form
$f(\theta^s|R) =\mathcal{N}(0,R^{-1})$, then the resulting EL
approximation to the marginal likelihood looks very much like the
Gaussian case, with the attending gains in efficiency if $C$ and $R$
share an exploitable structure, as in our derivation of the MPELE in
this model (recall section \ref{sec:comp-effic-maxim}).

This EL-approximate marginal likelihood has several potential
applications.  For example, marginal likelihoods are often used in the
context of model selection: if we need to choose between many possible
models indexed by the hyperparameter $R$, then a common approach is to
maximize the marginal likelihood as a function of $R$ \cite{gelman03a}.
Denote $\hat{R} = \arg \max_{R} \, \,F(R)$.  Computing $\hat R$
directly is often expensive, but computing the maximizer of the
EL-approximate marginal likelihood instead is often much cheaper.  In
our LNP example, for instance, the EL-approximate $\hat{R}$ can be
computed analytically if we can choose a basis such that $C =
\mathbf{I}$ and $R \propto \mathbf{I}$:
\begin{align}
\label{maxmargR}
 \hat{R} =
  \begin{cases}
    \frac{p}{\frac{q}{N_s^2} - \frac{p}{N_s}} \mathbf{I} & \text{if } p < \frac{q}{N_s} \\
   \infty       & \text{if } p \geq \frac{q}{N_s},
  \end{cases}
\end{align}
with $q\equiv\|X^Tr\|^2_2$ and $N_s = \sum_{n=1}^N r_n$; see methods
for the derivation.  Since $\marg = \Big(C + \frac{R}{\sum_n
  r_n}\Big)^{-1}\frac{X^Tr }{ \sum_n r_n}$ (equation \ref{regrsta}),
an infinite value of $R$ corresponds to $\marg$ equal to zero:
infinite penalization.  The intuitive interpretation of equation
\ref{maxmargR} is that the MPELE should be shrunk entirely to zero
when there isn't enough data, quantified by $\frac{q}{N_s}$, compared
to the dimensionality of the problem $p$.  Similar results hold when
there are correlations in the stimulus, $C\neq I$, as discussed in
more detail in the methods.

\begin{figure}[t!]
\begin{center}
\includegraphics[scale=0.99]{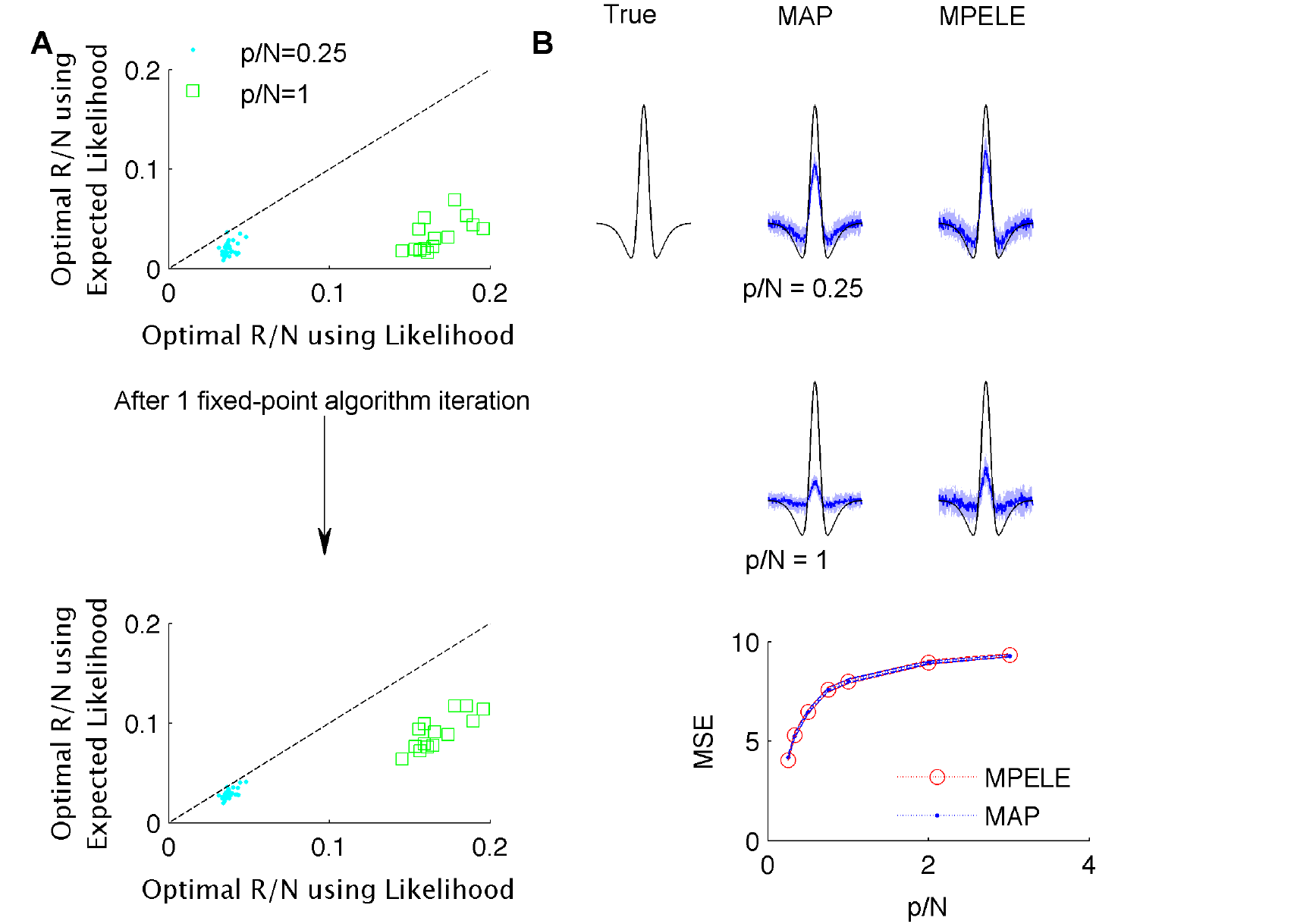}
\caption{The EL can be used for fast model selection via approximate
  maximum marginal likelihood.  Thirty simulated neural responses were
  drawn from a linear-nonlinear Poisson (LNP) model, with stimuli
  drawn from an independent white-noise Gaussian distribution.  The
  true filter (shown in black in B) has $p=250$ parameters and norm $10$.  A.) Optimal
  hyper-parameters $R$ (the precision of the Gaussian prior distribution) which maximize the marginal likelihood using
  the EL (top left column, vertical axis) scale similarly to those
  which maximize the full Laplace-approximated marginal likelihood
  (top left column, horizontal axis), but with a systematic downward
  bias.  After a single iteration of the fixed point algorithm used to
  maximize the full marginal likelihood (see text), the two sets of
  hyper-parameters (bottom left column) match to what turns out to be
  sufficient accuracy, as shown in (B): the median filter estimates (blue lines)
  ($\pm$ absolute median deviation (light blue), based on $30$ replications)
  computed using the exact and one-step approximate approach match for
  a wide range of $\frac{p}{N}$.  The MSE of the two approaches also
  matches for a wide range of $\frac{p}{N}$.}
\label{fig:evidence}
\end{center}
\end{figure}

Fig.~\ref{fig:evidence} presents a numerical illustration of the
resulting penalized estimators.  We simulated Poisson responses to
white-noise Gaussian stimuli using stimulus filters with $p=250$
parameters.  We use a standard iterative method for computing the
exact $\hat R$: it is known that the optimal $\hat{R}$ obeys
the equation 
\begin{eqnarray}
		\hat{R} &=& \frac{p - \hat{R} \sum_i H^{-1}(\map)_{ii}}{\map(\hat{R})^T \map(\hat{R})},
\end{eqnarray}
under the Laplace approximation \cite{bishop06a}.  Note that this equation
only implicitly solves for $\hat{R}$, since $\hat{R}$ is present on
both sides of the equation.  However, this leads to the following
common fixed-point iteration:
\begin{eqnarray}
\label{fpal}
		\hat R_{i+1} &=& \frac{p - \hat R_i \sum_i
                  H^{-1}(\map)_{ii}}{\map(\hat R_i)^T \map(\hat R_i)},
\end{eqnarray}
with $\hat R$ estimated as $\lim_{i \to \infty} \hat R_i$, when this
limit exists.  We find that the distance between the exact and
approximate $\hat{R}$ values increases with $\frac{p}{N}$, with the EL
systematically choosing lower values of $\hat{R}$ (Figure
\ref{fig:evidence}A top).  This difference shrinks after a single
iteration of equation \ref{fpal} initialized using equation
\ref{maxmargR} (Figure \ref{fig:evidence}A bottom).  The remaining
differences in $\hat{R}$ between the two methods did not lead to
differences in the corresponding estimated filters $\map(\hat R)$
(Figure \ref{fig:evidence}B).  In these simulations the exact MAP
typically took about 20 times longer to compute than a single
iteration of equation \ref{fpal} initialized using equation
\ref{maxmargR} (10 versus 0.5 seconds).

We close by noting that many alternative methods for model selection
in penalized GLMs have been studied; it seems likely that the EL
approximation could be useful in the context of other approaches based
on cross-validation or generalized cross-validation \cite{Golub79},
for example.  We leave this possibility for future work.  See
\cite{conroy12a} for a related recent discussion.

\subsection{Decreasing the computation time of Markov Chain Monte Carlo methods}
\label{sec:decr-comp-time}

As a final example, in this section we investigate the utility of the
EL approximation in the context of Markov chain Monte Carlo (MCMC)
methods \cite{robert05a} for sampling from the posterior distribution
of $\theta$ given GLM observations.  Two approaches suggest
themselves.  First, we could simply replace the likelihood with the
exponentiated EL, and sample from the resulting approximation to the
true posterior using our MCMC algorithm of choice \cite{sadeghi12a}.
This approach is fast but only provides samples from an approximation
to the posterior.  Alternatively, we could use the EL-approximate
posterior as a proposal density within an MCMC algorithm, and then use
the standard Metropolis-Hastings correction to obtain samples from the
exact posterior.  This approach, however, is slower, because we need
to compute the true log-likelihood with each Metropolis-Hastings
iteration.

\begin{figure}[t!]
\begin{center}
\includegraphics[scale=0.7]{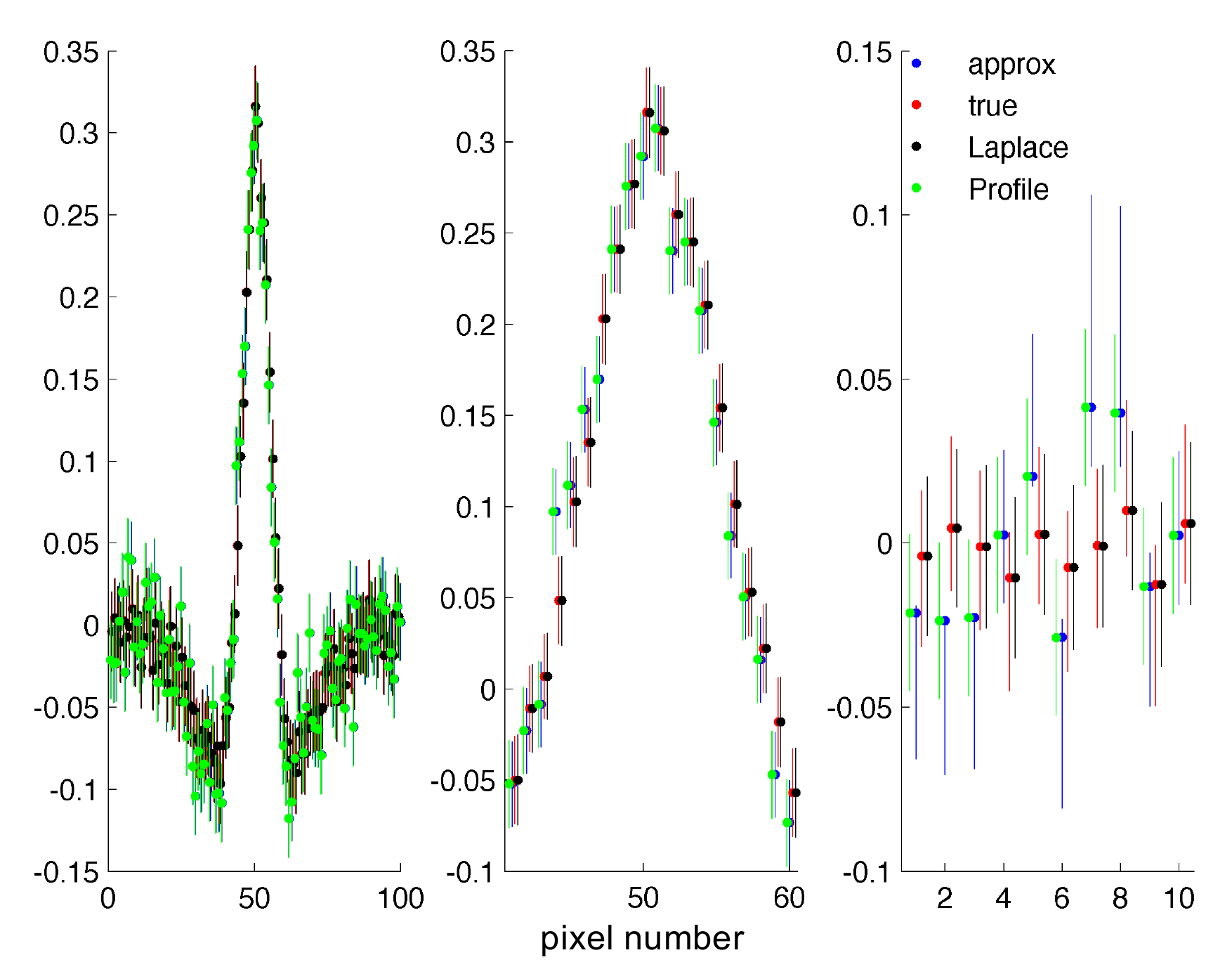}
\caption{The EL approximation leads to fast and (usually) accurate
  MCMC sampling from GLM posteriors.  $4000$ responses were simulated
  from a $100$-dimensional LNP model using i.i.d.\ white-noise
  Gaussian stimuli.  $10^6$ Markov Chain Monte Carlo samples, computed
  using Hybrid Monte Carlo, were then drawn from the posterior
  assuming a flat prior, using either the exact Poisson likelihood or
  the EL approximation to the likelihood.  The left column displays
  the median vector along with 95$\%$ credible regions for each
  marginal distribution (one marginal for each of the $100$ elements
  of $\theta$); approximate intervals are shown in blue, and exact
  intervals in red.  In the middle and right column we have zoomed in
  around different elements for visual clarity.  Statistics from both
  distributions are in close agreement for most, but not all, elements
  of $\theta$.  Replacing the EL-approximate likelihood with the
  EL-approximate profile likelihood yielded similar results (green).
  The Laplace approximation to the exact posterior also provided a
  good approximation to the exact posterior (black).  }
 \label{MCMCsamp}
\end{center}
\end{figure}

Figure \ref{MCMCsamp} illustrates an application of the first approach
to simulated data.  We use a standard Hamiltonian Monte Carlo
\cite{neal12a} method to sample from both the true and EL-approximate
posterior given $N=4000$ responses from a $100$-dimensional LNP model,
with a uniform (flat) prior.  We compare the marginal median and
95$\%$ credible interval computed by both methods, for each element of
$\theta$.  For most elements of the $\theta$ vector, the
EL-approximate posterior matches the true posterior well.  However, in
a few cases the true and approximate credible intervals differ
significantly; thus, it makes sense to use this method as a fast
exploratory tool, but perhaps not for conclusive analyses when $N/p$
is of moderate size.  (Of course, as $N/p \to \infty$, the EL
approximation becomes exact, while the true likelihood becomes
relatively more expensive, and so the EL approximation will become the
preferred approach in this limit.)

For comparison, we also compute the median and credible intervals
using two other approaches: (1) the standard Laplace approximation
(computed using the true likelihood), and (2), the profile EL-posterior,
which is exactly Gaussian in this case (recall section
\ref{sec:comp-effic-maxim}).  The intervals computed via (1) closely
match the MCMC output on the exact posterior, while the intervals
computed via (2) closely match the EL-approximate MCMC output; thus
the respective Gaussian approximations of the posterior appear to be
quite accurate for this example.

Experiments using the second approach described above (using
Metropolis-Hastings to obtain samples from the exact posterior) were
less successful.  The Hamiltonian Monte Carlo method is attractive
here, since we can in principle evaluate the EL-approximate posterior
cheaply many times along the Hamiltonian trajectory before having to
compute the (expensive) Metropolis-Hastings probability of accepting
the proposed trajectory.  However, we find (based on simulations
similar to those described above) that proposals based on the fast
EL-approximate approach are rejected at a much higher rate than
proposals generated using the exact posterior.  This lower acceptance
probability in turn implies that more iterations are required to
generate a sufficient number of accepted steps, sharply reducing the
computational advantage of the EL-based approach.  Again, as $N/p \to
\infty$, the EL approximation becomes exact, and the EL approach will
be preferred over exact MCMC methods --- but in this limit the Laplace
approximation will also be exact, obviating the need for expensive
MCMC approaches in the first place.

\section{Conclusion}

We have demonstrated the computational advantages of using the
expected log-likelihood (EL) to approximate the log-likelihood of a
generalized linear model with canonical link.  When making multiple
calls to the GLM likelihood (or its gradient and Hessian), the EL can
be computed approximately $O(Np/z)$ times faster, where $N$ is the
number of data samples collected, $p$ is the dimensionality of the
parameters and $z$ is the cost of a one-dimensional integral; in many
cases this integral can be evaluated analytically or
semi-analytically, making $z$ trivial.  In addition, in some cases the
EL can be analytically optimized or integrated out, making EL-based
approximations even more powerful.  We discussed applications to
maximum penalized likelihood-based estimators, model selection, and
MCMC sampling, but this list of applications is certainly far from
exhaustive.

Ideas related to the EL approximation have appeared previously in a
wide variety of contexts.  We have already discussed previous work in
neuroscience \cite{paninski04b,park11,field10a,sadeghi12a} that
exploits the EL approximation in the LNP model.  Similar
approximations have also been used to simplify the likelihood of GLM
models in the context of neural decoding \cite{rad11a}.  In the
Gaussian process regression literature, the well-known ``equivalent
kernel'' approximation can be seen as a version of the EL
approximation \cite{rasmussen05a,sollich05a}; similar approaches have
a long history in the spline literature \cite{Silverman84}.  Finally,
the EL approximation is somewhat reminiscent of the classical Fisher
scoring algorithm, in which the observed information matrix (the
Hessian of the negative log-likelihood) is replaced by the expected
information matrix (i.e., the expectation of the observed information
matrix, taken over the responses $r$) in the context of approximate
maximum likelihood estimation.  The major difference is that the EL
takes the expectation over the covariates $x$ instead of the responses
$r$.  A potential direction for future work would be to further
explore the relationships between these two expectation-based
approximations.

\clearpage

\section{Appendix: methods}
\subsection{Computing the mean-squared error for the MPELE and MAP in
  the linear-Gaussian model}

In this section we provide derivations for the results discussed in
section \ref{mathanaly}.  We consider the standard linear regression
model with Gaussian noise and a ridge (Gaussian) prior of the form
$f(\theta) \propto \exp \left( -cp\frac{\theta^T\theta}{2} \right)$,
with $c$ a scalar.  We further assume that the stimuli $x$ are
i.i.d.\ standard Gaussian vectors.  We derive the MSE of the MPELE and
MAP, then recover the non-regularized cases (i.e., the MLE and MELE)
by setting $c$ to zero.  Note that we allow the regularizer to scale
with the dimensionality of the problem, $p$, for reasons that will
become clear below.  The resulting MAP and MPELE are then found by
\begin{eqnarray}
    \map &=& \arg \max_\theta  \, \,  \theta^TX^Tr - \frac{1}{2}\theta^T (X^TX + cpI) \theta \\
    \label{amap} % analytical map
    	     &=& (X^TX + cpI)^{-1}X^Tr\\ 
    \marg &=& \arg \max_\theta  \, \,  \theta^TX^Tr - \frac{1}{2}(N + cp)\theta^T \theta\\
      \label{amarg} % analytical marg
    		  &=& \frac{X^Tr}{N + cp},
\end{eqnarray}
where we consider $X_{ij} \sim \mathcal{N}(0,1) \, \,  \forall i,j$.  For convenience of notation we define the quantity $\tilde{S} = X^TX + cpI$ and therefore write the MAP as $\map = \tilde{S}^{-1}X^Tr$.   

As usual the MSE can be written as the sum of a squared bias term and a variance term
\begin{eqnarray}
\label{bvdecomp}
          \mathbf{E}\Big[  \| \hat{\theta} - \theta \|^2_2 \Big] &=&  \| \epec{ (\hat{\theta} - \theta)} \|^2_2   + \mathbf{E}\Big[  \| \hat{\theta} - \mathbf{E}\Big[\hat{\theta} \Big]  \|^2_2 \Big].
\end{eqnarray}
The bias of the MAP equals
\begin{eqnarray}
  \epec{ (\map - \theta)} &=& \epec{\map} - \theta\\
  				     &=& \epec{\mathbf{E}\Big[\map|X\Big]} - \theta\\
				      &=& \epec{\tilde{S}^{-1}X^T \mathbf{E}\Big[r|X\Big]} - \theta\\
				     \label{cemap}
				     &=& \epec{\tilde{S}^{-1}X^TX\theta} - \theta,
\end{eqnarray}
The second line follows from the law of total expectation \cite{johnson07a} and the fourth follows from the fact $\mathbf{E}[r|X] = X\theta$.

From the law of total covariance, the variance can be written as
\begin{eqnarray}
\mathbf{E}\Big[  \| \map - \mathbf{E}\Big[\map \Big]  \|^2_2 \Big] &=& \mathbf{tr} \Big( \mathrm{Cov}(\map) \Big) \\
&=&  \mathbf{tr} \Big( \epec{\mathrm{Cov}(\map |X)} + \mathrm{Cov}(\mathbf{E}[\map|X])      \Big).
\end{eqnarray}
The term $\mathrm{Cov}(\map |X)$ equals
\begin{eqnarray}
\mathrm{Cov}(\map |X) &=& \tilde{S}^{-1}X^T\mathrm{Cov}(r|X)X\tilde{S}^{-1} \\
				    &=& \tilde{S}^{-1}X^TX\tilde{S}^{-1}.
\end{eqnarray}
In the second line we use the fact that $\mathrm{Cov}(r|X) = \mathbf{I}$.  The term $\mathbf{E}[\map|X]$ was used to derive equation \ref{cemap} and equals $\tilde{S}^{-1}X^TX\theta$.

Substituting the relevant quantities into equation \ref{bvdecomp}, we find that the mean squared error of the MAP is 
\begin{eqnarray}
\nonumber
\epec{ \| \map - \theta\|_2^2} &=&  \|  ( \epec{\tilde{S}^{-1}X^TX} - I)\theta \|^2_2 +\\
\label{intMAPe}
&& \mathbf{tr} \Big( \epec{\tilde{S}^{-1}X^TX\tilde{S}^{-1}} + \mathrm{Cov}(\tilde{S}^{-1}X^TX\theta)      \Big).
\end{eqnarray}

The MSE of the MPELE can be computed in a similar fashion.  The bias of the MPELE equals
\begin{eqnarray}
  \epec{ (\marg - \theta)} &=& \epec{\marg} - \theta\\
  				     &=& \epec{\mathbf{E}\Big[\marg|X\Big]} - \theta\\
				      &=& \epec{(N + cp)^{-1}X^T \mathbf{E}\Big[r|X\Big]} - \theta\\
				     \label{cemap2}
				     &=& (N + cp)^{-1}\epec{X^TX\theta} - \theta\\
				     &=& (N + cp)^{-1}N\theta - \theta.
\end{eqnarray}
To derive the fourth line we have again used the fact $\mathbf{E}[r|X] = X\theta$ to show that $\mathbf{E}[\marg|X] = (N + cp)^{-1}X^TX\theta$.  The fifth line follows by the definition $\epec{X^TX} = N\mathbf{I}$.  To compute the variance we again use the law of total covariance which requires the computation of a term $\epec{\mathrm{Cov}(\marg |X)}$,
\begin{eqnarray}
\epec{\mathrm{Cov}(\marg |X)} &=& \epec{(N + cp)^{-1}X^T\mathrm{Cov}(r|X)X(N + cp)^{-1} }\\
				    &=& \epec{(N + cp)^{-1}X^TX(N + cp)^{-1}}\\
				    &=& (N + cp)^{-2}N \mathbf{I}.
\end{eqnarray}
We use the fact that $\mathrm{Cov}(r|X) = \mathbf{I}$ to derive the second line and the definition $\epec{X^TX} = N\mathbf{I}$ to derive the third.
 
Using the bias-variance decomposition of the MSE, equation \ref{bvdecomp}, we find that the mean squared error of the MPELE estimator is 
\begin{eqnarray}
\epec{ \| \marg - \theta\|_2^2} &=&  \|  ( \frac{N}{N+cp} - 1)\theta \|^2_2 + \frac{1}{ (N+cp)^2} \Big(Np + \mathbf{tr} \Big( \mathrm{Cov}(X^TX\theta)      \Big) \Big).
\end{eqnarray}
The term $\mathbf{tr} \Big( \mathrm{Cov}(X^TX\theta)\Big)$ can be simplified by taking advantage of the fact that rows of $X$ are i.i.d normally distributed with mean zero, so their fourth central moment can be written as the sum of outer products of the second central moments \cite{johnson07a}.  
\begin{eqnarray}
	\epec{ (X^TX)_{ij} (X^TX)_{kl} }&=&  N \epec{X_{1i}X_{1j}X_{1k}X_{1l}} + N(N-1)\delta_{ij} \delta_{kl}\\
	&=&N(\delta_{ij}\delta_{kl} + \delta_{ik}\delta_{jl} + \delta_{il}\delta_{jk}) + N(N-1)\delta_{ij} \delta_{kl}.
\end{eqnarray}
We then have 
\begin{eqnarray}
\label{eqnint} % intermediate eqn
\epec{ \| \marg - \theta\|_2^2} &=&  \|  ( \frac{N}{N+cp} - 1)\theta \|^2_2 + \frac{1}{ (N+cp)^2} \Big(Np(1+ \| \theta\|^2_2) + N \| \theta\|^2_2    \Big).
\end{eqnarray}

Without regularization ($c=0$) the MSE expressions simplify significantly: 
\begin{eqnarray}
\epec{ \| \hat{\theta}_{MLE} - \theta\|_2^2} &=&  \mathbf{tr} \Big( \epec{(X^TX)^{-1}} \Big) \\
\epec{ \| \hat{\theta}_{MELE} - \theta\|_2^2} &=&   \frac{ \theta^T \theta + p( \theta^T \theta + 1) }{N}.
\end{eqnarray}
(The expression for the MSE of the MLE is of course quite well-known.)
Noting that $(X^TX)^{-1}$ is distributed according to an inverse
Wishart distribution with mean $\frac{1}{N-p-1}\mathbf{I}$
\cite{johnson07a}, we recover equations \ref{mmsea} and
\ref{mmse_MLa}.

For $c\neq0$ we calculate the MSE for both estimators in the limit $N,
p \rightarrow \infty$, with $0 < \frac{p}{N} = \rho < \infty$.  In
this limit equation \ref{eqnint} reduces to
\begin{eqnarray}
\label{margmap-asy}
  \epec{ \|\marg - \theta\|_2^2} &\rightarrow & \frac {\rho + 
  \theta^T\theta (c^2\rho^2 + \rho)} {(1+c\rho)^2}.
\end{eqnarray}

To calculate the limiting MSE value for the MAP we work in the eigenbasis of $X^TX$.  This allows us to take advantage of the Marchenko-Pastur law \cite{marchenko67a} which states that in the limit $N, p \rightarrow \infty$ but $\frac{p}{N}$ remains finite, the eigenvalues of $\frac{X^TX}{N}$ converge to a continuous random variable with known distribution.  We denote the matrix of eigenvectors of $X^TX$ by $O$:
\begin{eqnarray}
		X^TX &=& OLO^T,
\end{eqnarray}
with the diagonal matrix $L$ containing the eigenvalues of $X^TX$.

Evaluating the first and last term in the MAP MSE (equation \ref{intMAPe}) leads to the result
\begin{eqnarray}
\nonumber
 \|  ( \epec{\tilde{S}^{-1}X^TX} - I)\theta \|^2_2 + \mathbf{tr} \Big( \mathrm{Cov}(\tilde{S}^{-1}X^TX\theta)      \Big) &=& \| \theta \|^2_2  - 2\theta^T\epec{\tilde{S}^{-1}X^TX}\theta\\
 \label{dummyMAP}
 && + \epec{\|\tilde{S}^{-1}X^TX\theta\|^2_2}.
\end{eqnarray}
To evaluate the last term in the above equation, first note that
\begin{eqnarray}
\tilde{S}^{-1}X^TX\theta &=& O(L+cpI)^{-1}LO^T\theta.
\end{eqnarray}
Abbreviate $D = (L+cpI)^{-1}L$, for convenience.  Now we have
\begin{eqnarray}
\epec{\|\tilde{S}^{-1}X^TX\theta\|^2_2} &=& \theta^T\epec{O D^2
  O^T}\theta \\ &=& \epec{ \theta^T O D^2
  O^T \theta} \\ &=& \epec{ \mathbf{tr} (D^2
  O^T \theta \theta^T O) } \\ &=& \mathbf{tr} ~ \epec{ D^2
  O^T \theta \theta^T O } \\ &=& \mathbf{tr} ~ \epec{ D^2 \epec{
  O^T \theta \theta^T O \Big| D }} 
\end{eqnarray}
In the last line we have used the law of total expectation.  Since the
vector $O^T\theta$ is uniformly distributed on the sphere of radius
$\| \theta\|^2_2$ given $L$, $\mathbf{E}[O^T\theta\theta^TO|D] = \frac{\|
\theta\|_2^2}{p} \mathbf{I}$.  Thus
\begin{eqnarray}
\label{eq:dummyMAP2}
\epec{\|\tilde{S}^{-1}X^TX\theta\|^2_2} &=& \frac{ \| \theta\|_2^2}{p} \mathbf{tr}\Big( \epec{\Big((L+cpI)^{-1}L\Big)^2}\Big). 
\end{eqnarray}
We can use similar arguments to calculate the second term in equation
\ref{dummyMAP}.  
\begin{eqnarray}
\epec{\theta^T\tilde{S}^{-1}X^TX\theta} &=& \epec{
  \theta^TODO^T\theta}
\\
							    &=&  \mathbf{tr} ~ \epec{D}\frac{\|
\theta\|_2^2}{p}
\end{eqnarray}
Substituting this result and equation \ref{eq:dummyMAP2} into equation \ref{dummyMAP} we find
\begin{eqnarray}
 \|  ( \epec{\tilde{S}^{-1}X^TX} - I)\theta \|^2_2 + \mathbf{tr} \Big( \mathrm{Cov}(\tilde{S}^{-1}X^TX\theta)      \Big) &=&   \| \theta \|^2_2 - 2 ~ \mathbf{tr} ~ \epec{D}\frac{\|
\theta\|_2^2}{p} + \frac{ \| \theta\|_2^2}{p} \mathbf{tr} \epec{D^2}\\
 &=&  \frac{\| \theta\|_2^2}{p} \epec{ \mathbf{tr}(D -\mathbf{I})^2\Big)}\\
 &=&\frac{\| \theta\|_2^2}{p} \epec{ \mathbf{tr}\Big((L+cpI)^{-1}L -\mathbf{I})^2\Big)}.
\end{eqnarray}

Using the result given above and noting that $\mathbf{tr} \Big(
\epec{\tilde{S}^{-1}X^TX\tilde{S}^{-1}} \Big) = \mathbf{tr} \Big(
\epec{(L+cp\mathbf{I})^{-2}L}\Big) $, the MAP MSE can be written as
\begin{eqnarray}
\nonumber
\epec{ \| \map - \theta\|_2^2} &=& \mathbf{tr} \Big(\epec{ (L+cp\mathbf{I})^{-2}L}\Big) + \frac{\| \theta\|_2^2}{p} \epec{ \mathbf{tr}\Big((L+cpI)^{-1}L -\mathbf{I})^2\Big)}. \\
\end{eqnarray}
 
Taking the limit $N, p \rightarrow \infty$ with $\frac{p}{N}$ finite,
   \begin{eqnarray}
    \label{map-asy}
\epec{ \| \map - \theta\|_2^2} &\rightarrow& \rho\epec{\frac{l}{ (l +c\rho)^2}}  +  \| \theta \|_2^2\epec { \Big( \frac{l}{ (l +c\rho)} - 1\Big)^2},
\end{eqnarray}
where $l$ is a continuous random variable with probability density function $\frac{d\mu}{dl}$ found by the Marchecko-Pastur law
\begin{eqnarray}
		\label{mpd}
		\frac{d\mu}{dl} &=& \frac{1}{2 \pi l\rho} \sqrt{(b-l)(l-a)} \mathcal{I}_{[a, b]}(l)\\
		a\Big(\rho\Big) &=& (1-\sqrt{\rho})^2\\
		b\Big(\rho\Big) &=& (1+\sqrt{\rho})^2.
\end{eqnarray}
Using equation \ref{mpd} we can numerically evaluate the limiting MAP
MSE.  The results are plotted in figure \ref{toymse}.

\begin{figure}[t!]
\begin{center}
\includegraphics[scale=0.7]{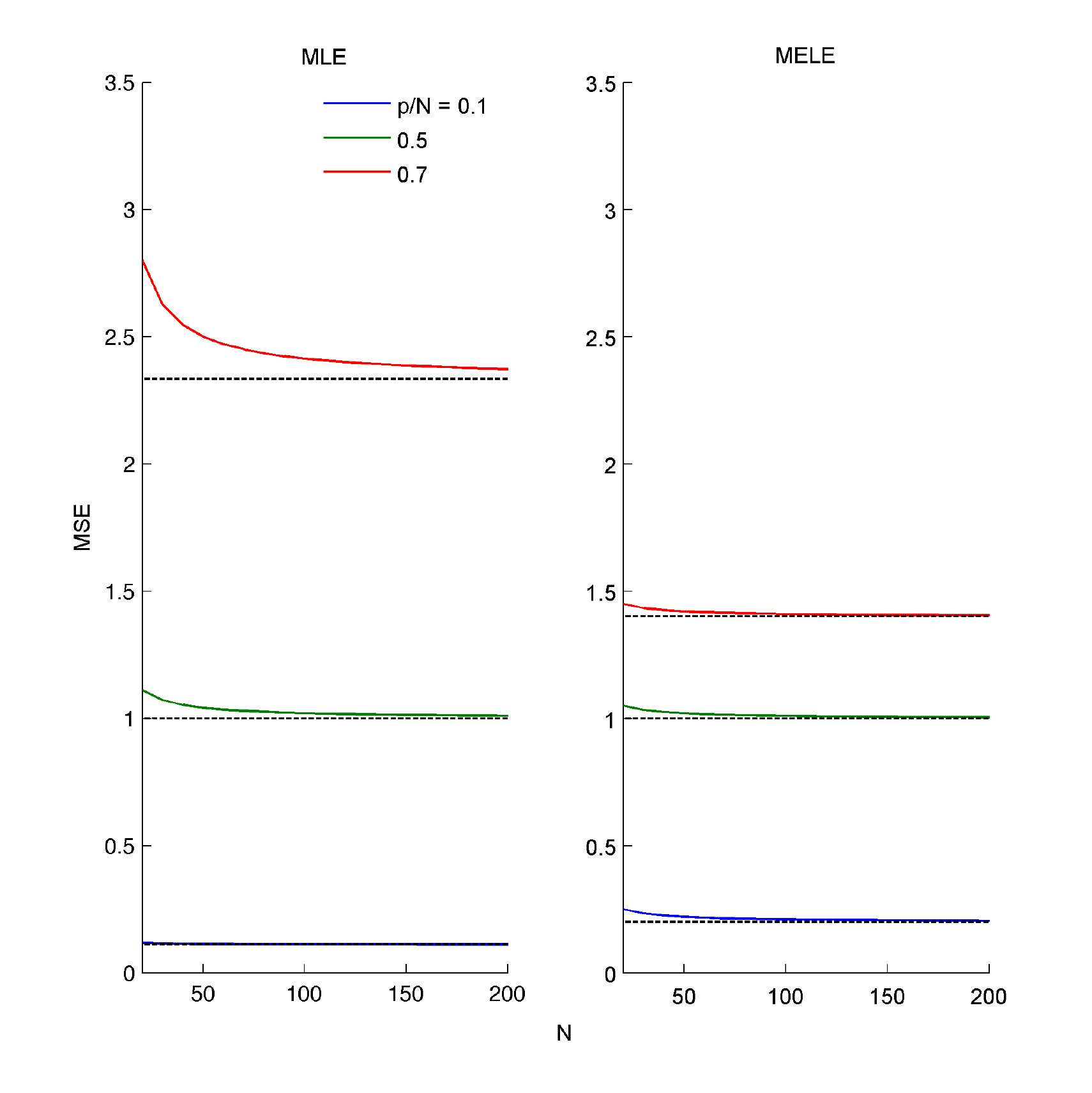}
\caption{Finite sample comparisons to the limiting values of the MLE
  and MELE MSE as $p,N \rightarrow \infty$, $\frac{p}{N} \rightarrow
  \rho$.  We plot the true mean-squared error (MSE) (colored lines) as
  a function of finite sample size for different values of
  $\frac{p}{N}$ for both the MLE (left column) and MELE (right
  column).  Black dashed lines show limiting MSE when $p,N \rightarrow
  \infty$, $\frac{p}{N} \rightarrow \rho$.  The quality of the
  approximation does not seem to depend on $\frac{p}{N}$ for the MELE
  and is within 1$\%$ accuracy after about 100 samples.  For the MLE,
  this approximation depends on $\frac{p}{N}$.  However, while the
  quality of the approximation depends on the estimator used, for data
  regimes common in neuroscience applications ($\frac{p}{N} \sim
  0.01$, $N\sim 10000$) the limiting approximation is acceptable.}
\label{finitesamp}
\end{center}
\end{figure}

Figure \ref{finitesamp} evaluates the accuracy of these limiting
approximations for finite $N$ and $p$.  For the range of $N$ and $p$
used in our real data analysis ($\frac{p}{N} \sim 0.01, N \sim
10000$), the approximation is valid.

\subsection{Computing the MPELE for an LNP model with Gaussian stimuli}
\label{MAPinhomo}

The MPELE is given by the solution of equation \ref{margMAP}, which in this case is
\begin{eqnarray}
\marg &=& \arg \max_{\theta, \theta_0} \, \, \theta_0\sum_{n=1}^Nr_n +
(X\theta)^Tr - N\epec{\exp(\theta_0)\exp(x^T\theta)} - \log
f(\theta).
\end{eqnarray}
Since $x\theta$ is Normally distributed, $ x\theta \sim
\mathcal{N}(0,\theta^T C\theta)$, we can analytically calculate the
expectation, yielding
\begin{eqnarray}
\label{mapint}
\marg &=& \arg \max_{\theta, \theta_0} \, \,  \theta_0\sum_{n=1}^Nr_n +  (X\theta)^Tr - N\exp(\theta_0)\exp \Big(\frac{\theta^TC\theta}{2}\Big)  - \log f(\theta).
\end{eqnarray}
Optimizing with respect to $\theta_0$ we find
\begin{eqnarray}
\label{moff}
		\exp( \theta_0^*) &=& \frac{\sum_{n=1}^Nr_n}{N}\exp(-\frac{\theta^TC\theta}{2}).
\end{eqnarray}
Inserting $ \theta_0^*$ into equation \ref{mapint} leaves the following quadratic optimization problem
\begin{eqnarray}
\label{profile}
\marg &=& \arg \max_{\theta} \, \, -\frac{\theta^T\Big(C
  \sum_{n=1}^Nr_n\Big)\theta}{2} + \theta^TX^Tr  - \log f(\theta).
\end{eqnarray}
Note that the first two terms here are quadratic; i.e., the
EL-approximate profile likelihood is Gaussian in this case.  If we use
a Gaussian prior, $f(\theta) \propto \exp(\theta^T R \theta /2)$, we
can optimize for $\marg$ analytically to obtain equation
\ref{regrsta}.

\subsection{Calculating $\hat{R}$ for the LNP model}
For the LNP model with $\epec{G(x^T\theta)}$ approximated as in equation \ref{eq:LNPapp} and $f(\theta|R) =\mathcal{N}(0,R^{-1})$, the Laplace approximation (\ref{lapp}) to the marginal likelihood yields 
\begin{eqnarray}
\nonumber
\log F(R)  &\approx&  \sum_{n=1}^N \hat{\theta}^*_0 r_n + (X\marg)^Tr -N\Delta t\exp(\hat{\theta}_0^*)\exp \Big(\frac{(\marg)^{T}C\marg}{2}\Big)  + \log\Big( f(\marg|R)\Big) \\
&& - \frac{1}{2} \log(\det(-H(\marg)))\\
%\log F(R)  &\approx&  \sum_{n=1}^N (\hat{\theta}^*)^T  r_n + (\hat{\theta}^*)^Tx_n^Tr_n - \epec{\sum_n G(x_n^T\hat{\theta}^*)} + \log\Big( f(\hat{\theta}^*|R)\Big)  - \frac{1}{2} \log(\det(-H(\hat{\theta}^*)))\\
\nonumber
	        &=&   -\frac{(\marg)^T C \marg}{2} \sum_{n=1}^Nr_n + (X\marg)^Tr   - \frac{(\marg)^T R\marg}{2} + \frac{1}{2}\log(\det(R))\\
	          && - \frac{1}{2} \log(\det(-H(\marg)))\\
	          \nonumber
	       &=&  -\frac{(\marg)^T(CN_s +R) \marg}{2} + (X\marg)^Tr  + \frac{1}{2}\log(\det(R))\\
	         \label{Rhatcalc1}
	         && - \frac{1}{2} \log(\det(-H(\marg))),
\end{eqnarray}
where the second line follows from substituting in equation \ref{moff} and we have denoted $\sum_{n=1}^Nr_n$ as $N_s$ in the third line.  Note that from the definition of the L2 regularized MPELE (equation \ref{regrsta})  we can write  
\begin{eqnarray}
X^Tr &=& (CN_s +R) \marg, 
\end{eqnarray}
and simplify the first two terms
\begin{eqnarray}
\nonumber
-\frac{(\marg)^T(CN_s +R) \marg}{2} + (X\marg)^Tr 	&=& \frac{(\marg)^T(CN_s +R) \marg}{2}\\
\label{simplefirsttwo}
										&=& \frac{(X^Tr)^T(CN_s +R)^{-1} X^Tr}{2}.
\end{eqnarray}
Noting that $C = \mathbf{I}, R = \beta\mathbf{I}$ equation \ref{simplefirsttwo} simplifies further to $ \frac{1}{2} \Big(N_s+ \beta\Big)^{-1}q$ with $q\equiv\|X^Tr\|^2_2$.  Using the fact that the profile Hessian is $-H(\marg) = CN_s+ R$ and the assumptions $C = \mathbf{I}, R = \beta\mathbf{I}$ the last two terms in equation \ref{Rhatcalc1} reduce to 
\begin{eqnarray}
\frac{1}{2}\log(\det(R))- \frac{1}{2} \log(\det(-H(\marg))) &=& \frac{p}{2}\log(\beta)- \frac{p}{2} \log(N_s + \beta).
\end{eqnarray}
Combining these results we find 
\begin{eqnarray}
\label{marglik_superapp}
\log F(R)  &\approx& \frac{1}{2} \Big(N_s+ \beta\Big)^{-1}q  + \frac{p}{2}\log(\beta)- \frac{p}{2} \log(N_s + \beta).
\end{eqnarray}
Taking the derivative of equation \ref{marglik_superapp} with respect
to $\beta$ we find that the critical points, $\hat{\beta}_c$ obey
\begin{eqnarray}
 0 &=& -\frac{q}{(N_s + \hat{\beta}_c)^2} + p\frac{N_s}{(N_s + \hat{\beta}_c)\hat{\beta}_c}, \\
  \hat{\beta}_c &=& \Big(   \frac{N_s^2p}{q - p N_s}, \infty \Big).
\end{eqnarray}
If $p \geq \frac{q}{N_s}$, the only critical point is $\infty$ since $\beta$ is constrained to be positive.  When $p < \frac{q}{N_s}$, the critical point $\hat{\beta}_c =   \frac{p}{\frac{q}{N_s^2} - \frac{p}{N_s}}$ is the maximum since $\log F$ (equation \ref{marglik_superapp}) evaluated at this point is greater than $\log F$ evaluated at $\infty$:
\begin{align}
	\log F \left(  \frac{N_s^2p}{q - p N_s} \right) = \frac{1}{2} \Big( \frac{q}{N_s} - p  + p\log\Big( \frac{p}{q}N_s\Big) \Big) \geq 0  = \lim_{\beta \to \infty} \log F. 
\end{align}
Therefore $\hat{R}$ satisfies equation \ref{maxmargR} in the text.  
  
  We can derive similar results for a more general case if $C$ and $R$ are diagonalized by the same basis.  If we denote this basis by $M$, we then have the property that the profile Hessian $CN_s+R = M(D^cN_s + D^r)M^T$ where $D^c$ and $D^r$ are diagonal matrices containing the eigenvalues of $C$ and $R$.  In this case the last two terms of equation \ref{Rhatcalc1} reduce to
\begin{eqnarray}
\frac{1}{2}\log(\det(R))- \frac{1}{2} \log(\det(-H(\marg))) &=& \frac{1}{2} \sum_{i=1}^p\log\Big( \frac{D^r_{ii}}{D^{c}_{ii}N_s +D^r_{ii}} \Big).
\end{eqnarray}
Defining $X^Tr$ rotated in the coordinate system specified by $M$ as $\tilde{q} = M^TX^Tr$, equation \ref{simplefirsttwo} simplifies to
\begin{eqnarray}
 \frac{(X^Tr)^T(CN_s +R)^{-1} X^Tr}{2} &=&  \frac{1}{2}\sum_{i=1}^p \tilde{q}_i^2(D^{c}_{ii}N_s +D^r_{ii})^{-1}.
\end{eqnarray}
Combining terms we find
\begin{eqnarray}
\log F(R)  &\approx& \frac{1}{2}\Big(\sum_{i=1}^p \tilde{q}_i^2(D^{c}_{ii}N_s +D^r_{ii})^{-1} + \log\Big( \frac{D^r_{ii}}{D^{c}_{ii}N_s +D^r_{ii}} \Big)\Big).
\end{eqnarray}
Taking the gradient of the above equation with respect to the eigenvalues of $R$, $D^{r*}_{jj}$, we find that the critical points obey
\begin{eqnarray}
0 &=& -\frac{\tilde{q}_j^2}{(D^c_{jj}N_s +D^{r*}_{jj})^2} +  \frac{D^c_{jj}N_s}{D^{r*}_{jj}(D^c_{jj}N_s +D^{r*}_{jj})} \\
D^{r*}_{jj} &=& \Big(   \frac{ (D^c_{jj}N^s)^2}{\tilde{q}_j ^2- D^c_{jj}N_s}, \infty \Big).
\end{eqnarray}
If $D^c_{jj}N_s \geq \tilde{q}_j^2 $, the only critical point is $\infty$ since $D^{r*}_{jj}$ is constrained to be positive (R is constrained to be positive definite).

\subsection{Real neuronal data details}

Stimuli are refreshed at a rate of 120 Hz and responses are binned at
this rate (figure \ref{LNPdata}) or at 10 times (figure \ref{glmnet})
this rate.  Stimulus receptive fields are fit with 81 (figure
\ref{LNPdata}) or 25 (figure \ref{glmnet}) spatial components and ten
temporal basis functions, giving a total of 81x10 = 810 or 25x10=250
stimulus filter parameters.  Five basis functions are delta functions
with peaks centered at the first 5 temporal lags while the remaining 5
are raised cosine `bump' functions \cite{pillow08a}.  The self-history
filter shown in figure \ref{glmnet} is parameterized by 4 cosine
`bump' functions and a refractory function that is negative for the
first stimulus time bin and zero otherwise.  The coupling coefficient
temporal components are modeled with a decaying exponential of the
form, $\exp{(-b\tau)}$, with $b$ set to a value which captures the
time-scale of cross-correlations seen in the data.  The errorbars of
the spike-history functions in figure \ref{glmnet}A show an estimate
of the variance of spike-history function estimates.  These are found
by first estimating the covariance of the spike-history basis
coefficients.  Since the L1 penalty is non-differentiable we estimate
this covariance matrix using the inverse log-likelihood Hessian of a
model without coupling terms, say $H_0^{-1}$, evaluated at the MAP and
MPELE solutions, which are found using the full model that
assumes non-zero coupling weights.  The covariance matrix of the
spike-history functions are then computed using the standard formula
for the covariance matrix of a linearly transformed variable.
Denoting the transformation matrix from spike-history coefficients to
spike-history functions as $B$, the covariance of spike-history
function estimates is $B^TH_0^{-1}B$.  Figure \ref{glmnet} plots
elements off the diagonal of this matrix.  We use the activity of
$100$ neighboring cells yielding a total of $100$ coupling coefficient
parameters, 5 self-history parameters, $250$ stimulus parameters, and
1 offset parameter ($356$ parameters in total).  The regularization
coefficients used in figure \ref{LNPdata}B and \ref{glmnet} are found
via cross-validation on a novel two minute (14,418 samples) data set.
Model performance is evaluated using 2 minutes of data not used for
determining model parameters or regularization coefficients.  To
report the log-likelihood in bits per second, we take the difference
of the log-likelihood under the model and log-likelihood under a
homogeneous Poisson process, divided by the total time.

The covariance of the correlated stimuli was spatiotemporally
separable, leading to a Kronecker form for $C$.  The temporal
covariance was given by a stationary AR(1) process; therefore this
component has a tridiagonal inverse \cite{paninski09a}.  The spatial
covariance was diagonal in the two-dimensional Fourier basis.  We were
therefore able to expoit fast Fourier and banded matrix techniques in
our computations involving $C$.

\subsection*{Acknowledgements}

We thank the Chichilnisky lab for kindly sharing their retinal data,
C.\ Ekanadham for help obtaining the data, and E.\ Pnevmatikakis,
A.\ Pakman, and W.\ Truccolo for helpful comments and discussions.  We
thank Columbia University Information Technology and the Office of the
Executive Vice President for Research for providing the computing
cluster used in this study.  LP is funded by a McKnight scholar award,
an NSF CAREER award, NEI grant EY018003, and by the Defense Advanced
Research Projects Agency (DARPA) MTO under the auspices of Dr.\ Jack
Judy, through the Space and Naval Warfare Systems Center, Pacific
Grant/Contract No.\ N66001-11-1-4205.

\bibliography{./MEbib}
\bibliographystyle{apalike}
\end{document}